\documentclass[twocolumn]{openjournal}
\usepackage{openjournal-submitted}
\usepackage{graphicx} 
\usepackage{multirow}

\usepackage[colorlinks,linkcolor=blue,citecolor=blue,urlcolor=blue ]{hyperref}
\usepackage[utf8]{inputenc}
\usepackage{float}
\usepackage{xcolor}
\usepackage{ulem}
\usepackage[T1]{fontenc}
\usepackage{amsmath}
\usepackage{xspace}
\usepackage{array,booktabs} 
\newcolumntype{P}[1]{>{\raggedright\arraybackslash}p{#1}} 

\newcommand{\enzo}{\textsc{Enzo~}}

\newcommand{\arepo}{\textsc{Arepo}}

\newcommand{\msolar}{$\rm{M_{\odot}~}$}
\newcommand{\msolarc}{$\rm{M_{\odot}}$}

\newcommand{\molH}{H$_{2}~$}
\newcommand{\smartstar}{\texttt{SmartStar~}}

\newcommand{\rarepeak} {\textsc{Rarepeak}\xspace}
\newcommand{\NormalOne} {\textsc{Normal1}\xspace}
\newcommand{\NormalTwo} {\textsc{Normal2}\xspace}

\defcitealias{Prole2023}{LP23}
\defcitealias{Schauer2021}{AS21}

\begin{document}


\title{The SEEDZ Simulations: Methodology and First Results on Massive Black Hole Seeding and Early Galaxy Growth\vspace{-1cm}}
\author{Lewis R. Prole$^{* \ 1}$}
\author{John A. Regan$^{1}$}
\author{Daxal Mehta$^{1}$}
\author{R\"udiger Pakmor$^{2}$}
\author{Sophie Koudmani$^{3,7}$}
\author{Martin A. Bourne$^{3,7,11}$}
\author{Simon C.~O. Glover$^{4}$}
\author{John H. Wise$^{5}$}
\author{Ralf S. Klessen$^{4,9}$}
\author{Michael Tremmel$^{6}$}
\author{Debora Sijacki$^{3,7}$}
\author{Ricarda S. Beckmann$^{10}$}
\author{Martin G. Haehnelt$^{3,7}$}
\author{John Brennan$^{1}$}
\author{Pelle van de Bor$^{1}$}
\author{Paul C. Clark$^{8}$}
\email{$^*$email: lewis.prole@mu.ie}
\affiliation{\\ \ \ \ \ $^{1}$ Centre for Astrophysics and Space Sciences Maynooth, Department of Physics, Maynooth University, Maynooth, Ireland.}
\affiliation{$^{2}$ Max-Planck-Institut f\"ur Astrophysik, Karl-Schwarzschild-Straße 1, D-85748, Garching, Germany}
\affiliation{$^{3}$ Institute of Astronomy, University of Cambridge, Madingley Road, Cambridge CB3 0HA, UK}
\affiliation {$^{4}$ Universit\"{a}t Heidelberg, Zentrum f\"{u}r Astronomie, Institut f\"{u}r Theoretische Astrophysik, Albert-Ueberle-Stra{\ss}e 2, 69120 Heidelberg, Germany.}
\affiliation {$^{5}$ Center for Relativistic Astrophysics, School of Physics, Georgia Institute of Technology, 837 State Street, Atlanta, GA 30332, USA}
\affiliation {$^{6}$ School of Physics, University College Cork, College Road, Cork T12 K8AF, Ireland}
\affiliation {$^{7}$ Kavli Institute of Cosmology, Cambridge, University of Cambridge, Madingley Road, Cambridge CB3 0HA, UK}
\affiliation {$^{8}$ Cardiff University School of Physics and Astronomy}
\affiliation{$^{9}$ Universit\"{a}t Heidelberg, Interdisziplin\"{a}res Zentrum f\"{u}r Wissenschaftliches Rechnen, Im Neuenheimer Feld 225, 69120 Heidelberg, Germany} 
\affiliation{$^{10}$ Institute for Astronomy, University of Edinburgh, Royal Observatory, Edinburgh EH9 3HJ, UK}
\affiliation{$^{11}$ Centre for Astrophysics Research, Department of Physics, Astronomy and Mathematics, University of Hertfordshire, College Lane, Hatfield, AL10 9AB, UK}

\begin{abstract}
\noindent Here we introduce the \texttt{SEEDZ} simulations, a large suite of cosmological hydrodynamic simulations exploring the formation and growth of the first black holes in the Universe, that may form through `light' or `heavy' seeding channels. Specifically, \texttt{SEEDZ} includes models for Population III star formation, supernova explosions and the resulting formation of `light' seed black holes. It also incorporates a model for subsequent metal enrichment, Population II star formation, `heavy' seed black hole formation, (super-)Eddington accretion as well as black hole feedback. In this paper, we cover the overall methodology employed and present our current results up to $z=15$. We find that black holes initially grow faster than their host galaxies, resulting in over-massive black holes with respect to the local black hole-galaxy scaling relations, when grown from seeds with $\sim$10$^4$ \msolar. At the high-end of our black hole masses, our simulated galaxy M$_{\rm BH}$ - M$_*$ relations match the observed high-redshift trends, i.e. over-massive black holes with M$_{\rm BH}$/M$_{\rm star} \sim 10^{-2}$. Hence, our simulations suggest that the fundamental black hole-galaxy relationships we observe at $z = 0$, especially the M$_{\rm BH}$ - M$_*$ relationship, likely only emerge in more mature galaxies. This initial set of simulations will be performed down to $z=10$, where we will conduct a comprehensive comparison of simulated black hole number densities and scaling relations with JWST observations. Further simulations with higher resolution will then follow.

\end{abstract}

\keywords{}

\maketitle
\section{Introduction} 
\label{sec:intro}
\noindent The fields of cosmology, high-redshift astrophysics, and gravitational-wave (GW) physics are entering a transformative era of discovery. The James Webb Space Telescope (JWST) is now unveiling a population of massive black holes (MBHs) at much earlier cosmic epochs than previously possible, extending deep into the high-redshift Universe and challenging long-standing assumptions about black hole and galaxy formation \citep{Maiolino2025, Matthee2024, Geris2025, Juodzbalis2025}. These detections provide direct evidence that MBHs with masses up to $10^6$ \msolar \citep{Maiolino2023} were already in place within the first few hundred million years of cosmic history, cementing long-standing questions about their seeding, growth, and the environments in which they formed. On the GW side, ESA’s formal adoption of the Laser Interferometer Space Antenna (LISA) represents an important step toward probing MBHs in a completely new observational window. Once launched in the mid-2030s, LISA will extend the detectable merger band beyond stellar-mass black holes into the MBH regime, enabling a direct census of their binary evolution and growth across cosmic time \citep{Amaro-Seoane2023}.

At present, electromagnetic (EM) observations are advancing at a remarkable pace, outstripping theoretical models and (unsurprisingly) repeatedly exposing shortcomings in our predictive frameworks. The unexpectedly high number densities of bright galaxies and MBHs revealed by JWST \citep{Harikane2023, Kocevski2023, Kocevski2024} require theoretical models that account for processes unique to the high redshift Universe (e.g. metal-free star formation, early black hole formation and feedback). It is within this context that the \texttt{SEEDZ} simulation suite has been developed. \texttt{SEEDZ} incorporates state-of-the-art prescriptions for Population III (PopIII) and Population II (PopII) star formation\footnote{When we refer to PopII stars we always mean PopII star clusters, as we do not resolve the stars within the cluster itself.}, the formation and evolution of light and heavy black hole seeds, and their subsequent growth and feedback. This paper introduces the first iteration of our \texttt{SEEDZ} model, with expected improvements in simulation resolution and modelling scheduled in the coming years.  

One of JWST’s most significant discoveries is the large population of $10^6$–$10^8$ \msolar\ MBHs existing within the first billion years of the Universe ($z > 6$) \citep[e.g.][]{Geris2025, Kocevski2023, Kocevski2024, Akins2024, Maiolino2023}. While their existence was anticipated from earlier quasar observations of $>10^8$ \msolar\ black holes \citep[e.g.][]{Banados2018, Matsuoka2019a, Wang2021, Yang2020, Fan2023}, the sheer number density of these less massive MBHs, coupled with the compactness of their host galaxies, has taken the community by surprise \citep[e.g.][]{Harikane2024, Matthee2024,Taylor2025a}. Explaining both the rapid growth of the MBHs and their prevalence in such compact systems has therefore become one of the most pressing challenges in modern astrophysics.

\indent Moreover, so-called "little red dot" (LRD) systems have been discovered. These systems, which are predominantly characterised by their compact morphology, V-shaped emission spectrum and red colour, are a source of intrigue within the community (e.g. \citealt{Labbe2023,Akins2024,Guia2024,Matthee2024, Perez-Gonzalez2024, Volonteri2024,Ananna2024,Baggen2024,Kocevski2025,Taylor2025a,Taylor2025}). While still hotly debated, it is entirely plausible that this new galaxy "type" hosts an active MBH at the centre with a MBH mass comparable to the entire stellar mass of the host \citep{Inayoshi2025, Juodzbalis2024, Juodzbalis2025, Durodola2024}. If true,  given that this population of galaxy represents a significant fraction of the overall galaxy population \citep{Matthee2023,Kocevski2023, Matthee2024, Kocevski2025} it has important implications for understanding the co-evolution of MBHs and galaxies. The \texttt{SEEDZ} simulation suite is well positioned to model this population as it incorporates the necessary MBH seeding pathways and galaxy formation requirements.

Theoretical models for MBH formation date back several decades, beginning with the seminal work of \cite{Rees1978}. Three mainstream seeding channels have been proposed: (i) remnants of PopIII stars \citep[e.g.][]{Madau2001}, (ii) the collapse of dense stellar or black hole clusters \citep[e.g.][]{Begelman1978,PortegiesZwart2004, Fragione2018}, and (iii) the formation of supermassive stars (SMSs) that collapse directly into black holes \citep{Woods2019} or `quasi-stars' \citep{Begelman2025}. While each pathway has been studied extensively, no one pathway has been definitively confirmed or indeed excluded, and multiple channels may well contribute simultaneously \citep{Regan2024a}. More speculative possibilities include primordial black holes \citep[e.g.][]{Hawking1971, Carr1974, Prole2025, Dayal2025} or exotic collapse mechanisms powered by dark matter annihilation \citep[e.g.][]{Freese2008, Spolyar2009, Tan2025}.

\indent Working our way through the main models, which we employ in \texttt{SEEDZ} and discuss in detail in \S \ref{sec:method}, we now briefly describe the rationale behind each seeding pathway. The most intuitive pathway through which MBHs may evolve is via the so-called light seed pathway. Light seeds are thought to form from the end points of PopIII stars and hence are expected to have initial masses in the range of a few solar masses up to several hundred solar masses, with a characteristic mass thought to be around 30 \msolarc \cite[e.g.][]{Hirano2014}. While at the first glance, light seeds appear to be an excellent vehicle to support MBH formation, they face a number of challenges and bottlenecks which have been highlighted in the literature beyond simply needing to grow by a large factor. These include initial hurdles to efficient growth \citep[e.g.][]{Smith2018}, the inability of light seed black holes to efficiently sink to the centre of the local potential \citep[e.g.][]{Pfister2019, Ma2021}, negative feedback from the black hole accretion process itself \citep{Milosavljevic2009, Alvarez2009} as well as negative feedback effects from nearby supernova explosions \citep[e.g.][]{Mehta2024}. Hence, while light seeds may potentially be highly ubiquitous, their ability to grow efficiently (by several orders of magnitude) to reproduce the observed MBH population is unclear. As a result, more exotic, but perhaps more viable pathways to MBHs have been explored.

\indent Heavy seed MBHs (here classed as black holes with seed masses in excess of 1000 \msolarc) can form either through the dense stellar/black hole cluster scenario or via the formation of a progenitor SMS. While both cases appear viable theoretically, they currently both lack a firm observational foundation. Moreover, more recent numerical models have shown that upper limits on the masses of seeds formed through both pathways are likely much lower than $10^5$ \msolar \citep{Regan2020, ArcaSedda2023a, Latif2022, Prole2024a, Rantala2025}. While these masses are clearly well in excess of the light seed masses, they are still far short of the masses of active black holes observed at high redshift by JWST. In fact, these heavy seeds still need to grow efficiently, perhaps even at super-Eddington rates, in order to reach the masses required, and negotiate similar hurdles that light seeds face (albeit starting from a larger seed mass).

\indent Our \texttt{SEEDZ} model incorporates both light and heavy seed formation pathways within a single cohesive framework. Rather than distinguish between SMS and star cluster-driven heavy seeds, we adopt criteria that stochastically generate heavy seeds where certain conditions are satisfied (see \S \ref{sec:method}), while remaining agnostic to the precise `sub-resolution' seeding channel. The layout of the paper is as follows: in \S \ref{sec:method} we introduce the numerical and physical methodology behind \texttt{SEEDZ}. In \S \ref{sec:box} we discuss the parent simulation box and subsequent zoom-in simulation regions. In \S \ref{sec:results} we outline the first results from our simulations before comparing our simulations with previous and current generation simulations in \S \ref{sec:discussion}. Finally, in \S \ref{sec:conclusions} we summarise our results within the context of the current observational landscape.

\section{Numerical method}
\label{sec:method}
The \texttt{SEEDZ} simulations are run using the state-of-the-art moving mesh code \textsc{arepo2} \citep{Springel2010, Pakmor2016, Pakmor2023}, which solves the Euler equations on an
unstructured Voronoi mesh. The fluid dynamics are solved using a finite-volume, second-order reconstruction scheme coupled with an exact Riemann solver. The unstructured mesh adapts by locally refining and de-refining in a quasi-Lagrangian fashion. We employ the hierarchical time integration scheme described in \cite{Springel2021}, which splits the Hamiltonian, for gravity only, into ``fast'' and ``slow'' components, i.e. those on short versus long time-steps, respectively. With this approach, interactions between particles/cells on the smallest time steps (up to a maximum of 30000 particles) are solved using direct summation, while a standard octree approach is used for all other interactions. This is primarily done for efficiency, to avoid unnecessary tree constructions for small numbers of particles, but has the added benefit of more accurately calculating the gravitational forces between black holes and their nearest neighbours, which are on the shortest time-steps, via direct summation \citep[see e.g.][]{Bourne2024}.

\indent Below the gas number density threshold of 1~cm$^{-3}$, the default `Lagrangian' refinement/de-refinement \arepo~ strategy is employed, whereby gas cell mass is within the factor of two of the target baryonic mass resolution value. Above the gas number density threshold of 1~cm$^{-3}$, refinement of the mesh is triggered by the Jeans refinement criterion, with cells refining such that the Jeans length must be resolved by at least 4 cells. We now describe the sink particle implementation, the accretion and feedback sub-grid models employed, the chemistry solver used and finally the boxsize, softening lengths and zoom-in configurations used. 

\subsection{Seeding Stellar and Black Hole Masses} \label{Sec:seeding}
\noindent \texttt{SEEDZ} models stars and black holes as sink particles \citep{Bate1995}, building on the implementation from \cite{Tress2020} and \cite{Schauer2021}. The formation of a sink particle is triggered when gravitational instability occurs in a cell that has refined down to our maximum resolution, i.e. when the Jeans length of a cell becomes smaller than the minimum allowed cell size (62 cpc h$^{-1}$ or 5.8 physical pc at $z=15$), following on from the initial implementation of this methodology in \cite{Krumholz2004} and \cite{Krumholz2006}. We define a sink particle formation region surrounding the candidate cell $R_{\rm f}$, with a radius of 5 times the minimum cell size. Sink particles can only form when the following criteria are met:
\begin{itemize}\setlength\itemsep{0.2em} 
    \item The host cell is at the highest refinement level, such that the cell volume is within a factor of 2 of the set minimum volume parameter
    \item The cell density exceeds the Jeans density as set by the minimum cell size and the cell's current temperature
    \item The cell velocities and accelerations within the formation region are converging towards the host cell
    \item The host cell is not within the accretion region of another sink particle
\end{itemize}
Once the above four checks have been satisfied, a sink particle is formed. 
At this point a second set of checks are initiated to assess the appropriate treatment.  Our sink particle formation framework, which we dub \smartstar \citep{Regan2020, Regan2018, Regan2018a}, and which has previously been used in a modified form in the \enzo code \citep{Enzo-Collaboration2014, Brummel-Smith2019} allows three distinct initial sink particle outcomes based on the properties of the formation environment and the host cell. They are as follows and are checked in order:
\begin{itemize}
  \item The initial accretion rate onto the host cell is estimated by summing radial mass flux contributions from all cells within the formation environment, $R_{\rm f}$. If the mass flux is calculated to be above a threshold of 1 M$_{\odot}$ yr$^{-1}$, a MBH is formed directly \citep{Regan2023a}. We employ no metallicity or \molH threshold. SMS formation does not depend on the \molH fraction \cite[e.g.][]{Woods2019} and while a highly metal enriched region will fragment into a star cluster, this may also lead to the formation of a heavy seed given the extreme accretion and mass budget \citep[e.g.][]{ArcaSedda2023, Rantala2025}. The initial mass of the MBH is extracted from a power-law functional form starting at $5 \times 10^3$ \msolar and ending at $10^5$ \msolarc. The functional form is a simple power law with a slope of -0.5. We are agnostic, at our current resolution, to the physical processes leading to the MBH (be it SMS formation or the collapse within a dense stellar or black hole cluster). In cases where the host cell contains insufficient mass to form a SMS of the "correct" mass (as determined by our IMF) we instead take 90\% of the mass of the host cell instead and use that as the SMS mass. At our current resolution (cell size) this is rarely, if ever, an issue. 

    \item If the initial mass flux onto the host cell from the formation environment is less than 1 M$_{\odot}$ yr$^{-1}$ and the host cell has a metallicity greater than $10^{-4}$ Z$_\odot$, its mass is converted into a PopII stellar cluster particle. The stellar population hosted by the PopII stellar cluster is sampled from a Kroupa IMF \citep{kroupa2001, Clark2008, Dopcke2011, Dopcke2013}. At our resolution (see \S \ref{sec:box}), the PopII cluster particles are typically 10$^4$-10$^5$ \msolarc, which allows sufficient sampling of the IMF. Stars between 8 \msolar and 104 \msolar end their lives as supernova (SN) explosions or direct collapse black holes (see \S \ref{sec:SNe}), while stars sampled below 8 \msolar remain in the cluster throughout the simulation (i.e. that mass is not returned to the ISM). 
    
    \item Finally, when gravitational instability is triggered in a potential host cell with a metallicity below the threshold of $10^{-4}$ Z$_\odot$ (see Section \ref{sec:SNe}) and a H$_2$ fraction greater than $5 \times 10^{-4}$, a singular PopIII star particle is formed. The stellar mass of the PopIII star is randomly sampled from a top-heavy M$^{-1.3}$ initial mass function (IMF) in the range of 1-300 M$_{\odot}$ \citep[][]{Wise2011, Chen2014a, Prole2023}. The corresponding mass is removed from the host gas cell. Depending on their mass,
    PopIII stars can end their lives in one of three ways: via an SN explosion resulting in complete disruption of the star and the deletion of the star particle, via an SN that forms a light seed BH particle, or by direct collapse to form a light seed BH. See the next subsection for more details. 
\end{itemize}

To prevent numerically artificial and instantaneous mergers between BHs, we do not allow BH or PopIII stars (which soon become BHs) to form within the accretion radius of a pre-existing BH or PopIII star. However, since PopII clusters do not merge (see \S \ref{sec:mergers}) we do not include an exclusion zone around PopII cluster particles. In this way, the stellar mass content of halos is not artificially reduced.

\subsection{SN Explosions and Metal Enrichment}
\label{sec:SNe}
\noindent Our SN treatment builds on our initial implementation in \cite{Mehta2024}, where we initially tested the \texttt{SEEDZ} subgrid model. Upon being assigned a stellar mass in the case of PopIII star formation, or multiple stellar masses in the case of Pop II cluster formation, stars are also assigned a stellar lifetime using Table 25.6 of \cite{Maeder2009}. When the stellar lifetime has been exceeded, the particle type and mass is checked and one of five different scenarios is invoked: 
   \begin{itemize}
      \item Stars in the mass range 11-20 M$_\odot$ undergo a core collapse SN, with energy $E_{\rm SN} = 10^{51} $ erg ($E_{51} = 1$). In the case of PopIII stars, the particle transitions into a BH particle following the explosion. In the case of a PopII cluster the explosion mass ($m_{\rm ex} = m_{\star} - m_{\rm He}$) is simply subtracted from the cluster and the mass returned to the grid.  The BH mass remaining after the explosion is the helium core mass, calculated as \citep{Nomoto2006}
      \begin{equation}
    m_{\rm He} = 0.1077 +  0.3383 (m_{\star}-11) \: {\rm M}_{\odot},
    \label{eq:He_core_small}
    \end{equation} 
    where $m_{\star}$ is the stellar mass in units of M$_\odot$. For PopIII stars, the BH remnant is a light seed black hole (the remnants left behind by PopII stars within a PopII star cluster would in reality also be light seeds. Our current modelling does not track accretion onto those seeds). 
    
    \item Stars in the mass range of 20-40 M$_\odot$ undergo a hypernova. We extrapolate the values of $E_{\rm{SN}}$ and m$_{\rm He}$ from \citet{Nomoto2006}. Following the explosion, PopIII star particles again transition into BH particles with their mass set to the helium core mass post explosion. For PopII stellar clusters the explosion mass is removed from the cluster and returned to the grid. 
    
    \item Stars in the mass range of $40-140$ M$_\odot$ experience a direct collapse into a BH and exhibit no SN explosion. The total mass of the PopIII particle is converted into a BH particle. For PopII clusters nothing happens. 
    
    \item Between 140 - 260 M$_\odot$, PopIII stars undergo a pair-instability SN, with their values of $E_{\rm{SN}}$ and m$_{\rm He}$ calculated as
      \begin{equation}
      m_{\rm He} = \frac{13}{24} (m_{\star} - 20)
      \label{eq:He_core_large}
      \end{equation}
      \begin{equation}
      E_{\rm SN} = \left[5 + 1.304 (m_{\rm He} - 64) \right] \times 10^{51} \: {\rm erg}.
      \label{eq:SN_E51}
      \end{equation}
    These particles are not converted to BHs after the explosion and are instead deleted from the simulation.
    
    \item Finally, stars with masses above 260 M$_\odot$ also fall into the direct collapse category and PopIII stars are converted into BH particles without an SN explosion.
    \end{itemize}

As the mass of PopII cluster particles is significantly greater than that of PopIII particles, we return the ejecta mass of the star (the difference between the stellar and helium core mass) to the injection region (defined below) after each explosion. Once the final SN assigned to a PopII cluster particle has occurred, the mass of the cluster can no longer be depleted and the remaining PopII particle represents a collection of low mass stars.

As in \cite{Magg2022}, we model SN explosions by injecting momentum into the surrounding cells. We use  a fixed explosion region radius of 10 times the minimum cell diameter ($\sim$ 150 pc). With our refinement scheme, this typically  contains $\sim$ 1000 cells. We calculate the momentum of the blast according to the SN energy, density and metallicity of the injection region as in \cite{Smith2019}. 
\begin{equation}
    p_{\rm SN} = 3.0 \times 10^5 E_{\rm 51}^{16/17} n_{\rm SN}^{-2/17} Z_{\rm SN}^{-0.14} \, {\rm M}_\odot \rm{km/s},
    \label{eq:mom_inj}
\end{equation}
where $E_{\rm 51} = E_{\rm SN}/10^{51} \: {\rm erg}$, $n_{\rm SN}$ is the mass weighted number density of the injection region, and $Z_{\rm SN}$ is the mass weighted metallicity in units of solar metallicity. As the metallicity of the injection region can be 0 in some regions, especially in the early Universe, we use the maximum value between $Z_{\rm SN}$ and 0.01. Finally, we fully ionise the gas within the injection region by simply setting the chemical abundances to 100$\%$ ionized hydrogen. We then update the total energy of the cells to reflect the new kinetic and internal energies of the gas. 

After each explosion, the mass of metals produced is calculated as the difference between the initial stellar mass and the final BH (helium core) mass. We model the injection and subsequent mixing of these metals as a passive tracer field, initially assigned to the cells in the injection region, with each cell receiving a fraction proportional to their mass. As the mass from these cells flows through the cell boundaries into their surrounding cells, the metal tracer field also advects a proportional amount of the metal mass into those cells. The metallicity of a cell is calculated as its metal mass divided by the regular gas mass of the cell. It is important to note that the metal tracer field does not add any mass to the mesh, but acts as its own independent tracer field in the simulation.

\subsection{Black Hole Accretion}
\noindent Upon formation, PopIII stars and BH particles are assigned an accretion radius $R_{\rm acc}$ set to 5 times the global minimum cell length. This accretion scale is a co-moving value, which grows throughout the simulation with the expansion of the Universe. While only BH particles can accrete gas from their surroundings, PopIII stars retain their accretion radii after they are converted into BH particles and can begin accreting. PopII cluster particles are not assigned an accretion radius, such that they are ignored during merger (\S \ref{sec:mergers}) and formation (\S \ref{Sec:seeding}) checks.

We model accretion onto BHs as Bondi-Hoyle-Lyttleton \citep{Bondi1952} accretion, again following the formulation set out in \cite{Krumholz2004} and \cite{Krumholz2006}. The Bondi radius is given by
\begin{equation}
        R_{\rm Bondi} = \frac{G M_{BH}}{v_{\infty}^2+c_{\infty}^2}\,,
        \label{eq:bondi_radius}
\end{equation}
where $v_{\infty}$ is the mass-weighted speed of the gas within the accretion radius relative to the BH and $c_{\infty}$ is the sound speed in the region. 
The accretion rate onto the BH is then calculated using the usual Bondi formula
\begin{equation}
    \dot{M}_{\rm Bondi} = 4 \pi \rho_{\infty} R_{\rm Bondi}^2 /(1.12 c_{\infty})^2 + v_{\infty}^2)^{1/2}\,,
    \label{eq:bondi-hoyle-acc}
\end{equation}
where $\rho_{\infty}$ is the weighted density for each cell inside the accretion radius, computed as described below.

Since the Bondi radius scales with the square of the black hole mass, we cannot guarantee that we can always resolve the Bondi radius;
indeed, for lower mass light seed black holes, we do not resolve their Bondi radius in any of the simulations presented in this paper. In order to account for this, we follow \cite{Krumholz2004} and define a kernel radius which accounts for the resolution of the simulation relative to the Bondi radius. The kernel radius, r$_{\rm K}$, is computed as:
\begin{equation}
r_{\rm K} = \left\{ \begin{array}{lcr}
  \Delta x_{\rm min}/4 & &R_{\rm Bondi} <  \Delta x_{\rm min}/4\\
  R_{\rm Bondi}  & & \ \ \ \ \Delta x_{\rm min}/4 \le R_{\rm Bondi} \le R_{\rm acc}/2\\
  R_{\rm acc}/2 && R_{\rm Bondi} > R_{\rm acc}/2
\end{array} \right.
\end{equation}
where $\Delta x_{\rm min}$ is the current minimum cell length. The kernel radius is used to assign a weight to every cell within $R_{\rm acc}$ using
\begin{equation}
    W \propto \exp(-r^2/r_{\rm K}^2),
\label{eq:weights}
\end{equation}
where $r$ is the distance from the cell to the accreting BH. The weighted density $\rho_{\infty}$ (Equation \ref{eq:bondi-hoyle-acc}) is then calculated as:
\begin{equation}
     \rho_{\infty} = \bar{\rho}  W
\end{equation}
where $\bar{\rho}$ is the mass-weighted mean density within the accretion sphere.

\subsubsection{Vorticity Adjustment}
\noindent Following on from \cite{Krumholz2006} and \cite{Mehta2024}, we further adjust the accretion rate based on the vorticity $\omega$ of the surrounding gas, given by
\begin{equation}
    \omega = | \nabla \times \vec{v} |
\end{equation}
with the dimensionless vorticity $\omega_*$ given by
\begin{equation}
\omega_* =  \omega  \frac{R_{\rm Bondi}}{c_{\infty}}.
\end{equation}
We introduce a damping factor $f(\omega)$ defined as
\begin{equation}
    f_{w} = \frac{1}{1 + \omega_*^{0.9}}
\end{equation}
and calculate the accretion rate in a turbulent medium according to
\begin{equation}
    \dot{M}_{\omega} = 4  \pi  \rho_{\infty}  R_{\rm Bondi}^2  c_{\infty}  (0.34  f_{\omega_*}).
\end{equation}
The total accretion rate onto the MBH particle is
\begin{equation}
    \dot{M} = (\dot{M}_{\rm Bondi}^{-2} + \dot{M}_{\omega}^{-2})^{-0.5},
\end{equation}

For a given time step, $t_h$, the mass of the MBH particle increases by $M_{\rm acc} = t_h \dot{M}$, which is removed from cells within $R_{\rm acc}$ using the weighting scheme calculated in Equation \ref{eq:weights}, adjusting the velocities of both the gas cells and the BH to conserve linear momentum. The BH position is then shifted to the center of mass of a system comprised of the BH and the accreted mass contributions from each cell.

\subsection{Black Hole Accretion Feedback}
\noindent To account for BH accretion feedback, we inject thermal energy into gas cells within the accretion region isotropically based on the accretion rate onto a BH. The total mass accreted during a single timestep is given by $M_{\rm acc}$. This produces an amount of thermal energy
\begin{equation}
    E_{\rm th, acc} = \epsilon f_c  M_{\rm acc}c^2,
\end{equation}
where $f_c$ is the thermal coupling factor, which we set to 0.05 \citep[see e.g.][]{Booth2009}, and $\epsilon$ is the radiative efficiency. For sub-Eddington accretion rates, we calculate $\epsilon$ as
\begin{equation}
    \epsilon = 1 - \sqrt{1-\frac{2}{3 R_{\rm ISCO}}},
\label{eq:epsilon}
\end{equation}
where $R_{\rm ISCO}$ is the innermost stable orbit. Assuming the BH's spin is prograde with respect to the accretion disc, $R_{\rm ISCO}$ is given by
\begin{equation}
    R_{\rm ISCO} = 3 + r_2 - \sqrt{(3 - r_1) * (3 + r_1 + 2 * r_2)},
\end{equation}
in units of $GM/c^2$ and $r_1$ and $r_2$ are given by
\begin{align}
r_1 &= 1 + (1 - a^2)^{1/3}  (1 + a)^{1/3} + (1 - a)^{1/3}, \\
r_2 &= \sqrt{3  a^2 + r_1^2},
\end{align}
and $a$ is the dimensionless black hole spin, assumed to be 0.7. These assumptions give a radiative efficiency, $\epsilon$ $\sim$ 0.1.

The calculation of $\epsilon$ changes for super-Eddington accretion rates, where the Eddington limit is
\begin{equation}
    \dot{M}_{\rm Edd}=\frac{4 \pi G M m_p}{ \epsilon c \sigma_{\rm T}},
\end{equation}
where $G$ is the gravitational constant, $c$ is the speed of light, $\epsilon$ is the sub-Eddington value calculated using Equation \ref{eq:epsilon} and $\sigma_{\rm T} = 6.65 \times 10^{-25} \: {\rm cm^{2}}$ is the Thomson scattering cross-section for an electron. In this case, we first find the super-Eddington luminosity by calculating the Eddington luminosity as
\begin{equation}
    L_{\rm Edd}=\frac{4 \pi G M m_p \mu c}{\sigma_{\rm T}},
\end{equation}
where $\mu$ is the mean molecular weight, assumed to be 1.22. We then use the parametrisation found in \citet{Madau2014} to calculate the super-Eddington luminosity
\begin{equation}
    L_{\rm SE}=L_{\rm Edd} A \left(  \frac{0.985}{r_{\rm Edd} + B} + \frac{0.015}{r_{\rm Edd} + C} \right),
\end{equation}
where $r_{\rm Edd}$ is the ratio of of the accretion rate to the Eddington limit, and
\begin{equation}
\begin{aligned}
A =& (0.9663 - 0.9292  a)^{-0.5639}, \\
B =& (4.627 - 4.445  a)^{-0.5524},\\
C =& (827.3 - 718.1  a)^{-0.7060},\\
\end{aligned}
\end{equation}
where again the BH spin $a$ is set to 0.7. The value of $\epsilon$ in the super-Eddington case then follows as
\begin{equation}
\epsilon  = L_{\rm SE} / (\dot{M} c^2).
\end{equation}
We use this updated value of the radiative efficiency in our super-Eddington calculation. 
Note that in reality super-Eddington accretion will also drive mechanical feedback \citep[see e.g.][]{Regan2019}. We do not employ mechanical (jet) feedback here but note that it is likely to be a signature of super-Eddington accretion. 

\subsection{Chemistry Solver}
\noindent Chemistry and cooling are modelled using a state-of-the-art non-equilibrium chemistry solver which tracks hydrogen, deuterium, 
helium, molecular hydrogen and their
corresponding ions using the primordial chemistry network from \cite{clark2011}, with some updated rate coefficients introduced in \cite{schauer2017}.

The effects of metal cooling are not included in this study as they are generally 
not relevant at the low metallicities and densities we reach in our simulation \citep{Jappsen2007, Smith2015, Chiaki2016}. As noted above, however, a metallicity tracer field is implemented allowing us to track the diffusion of metal products in our simulations. Radiative transfer is not included in this simulation suite.

\subsection{Lyman-Werner Background Evolution}
\noindent As the Universe evolves and the first generation of PopIII and PopII stars builds throughout the Universe, photons from these stars are expected to form a ubiquitous background radiation field which can alter the chemistry of the ISM and IGM \citep{Haiman2000}. In particular, photons in the Lyman-Werner (LW) band can photodissociate H$_2$ molecules, the most important coolant for collapse and star formation in the early Universe. Reducing the H$_2$ content of early halos is thought to delay star formation in 10$^6$ M$_\odot$ minihalos \citep[e.g.][]{OShea2008}.

\indent As we have not implemented radiative transfer from our star particles, it is necessary to include a uniform background radiation field characterized by its intensity in the LW band, $\rm{J_{LW}}$, in units of J$_{21}$ = 10$^{-21}$ erg s$^{-1}$ Hz$^{-1}$ sr$^{-1}$ cm$^{-2}$, which is used within the chemical network to modify the rates of chemical reactions appropriately to account for the missing radiation. As originally implemented in \cite{Prole2025}, we extrapolate the background LW intensity from the reference simulation of \cite{Qin2020} presented in their Figure 5, which begins at $z \sim 30$ at very low values of $\rm{J_{LW}}$ $\sim 10^{-3}$ J$_{21}$ and climbs to $\rm{J_{LW}}$ $\sim$ 2 J$_{21}$ by $z \sim 10$.
\begin{figure*}
  \centering
  \hbox{\hspace{3cm}
  \includegraphics[width=13cm]{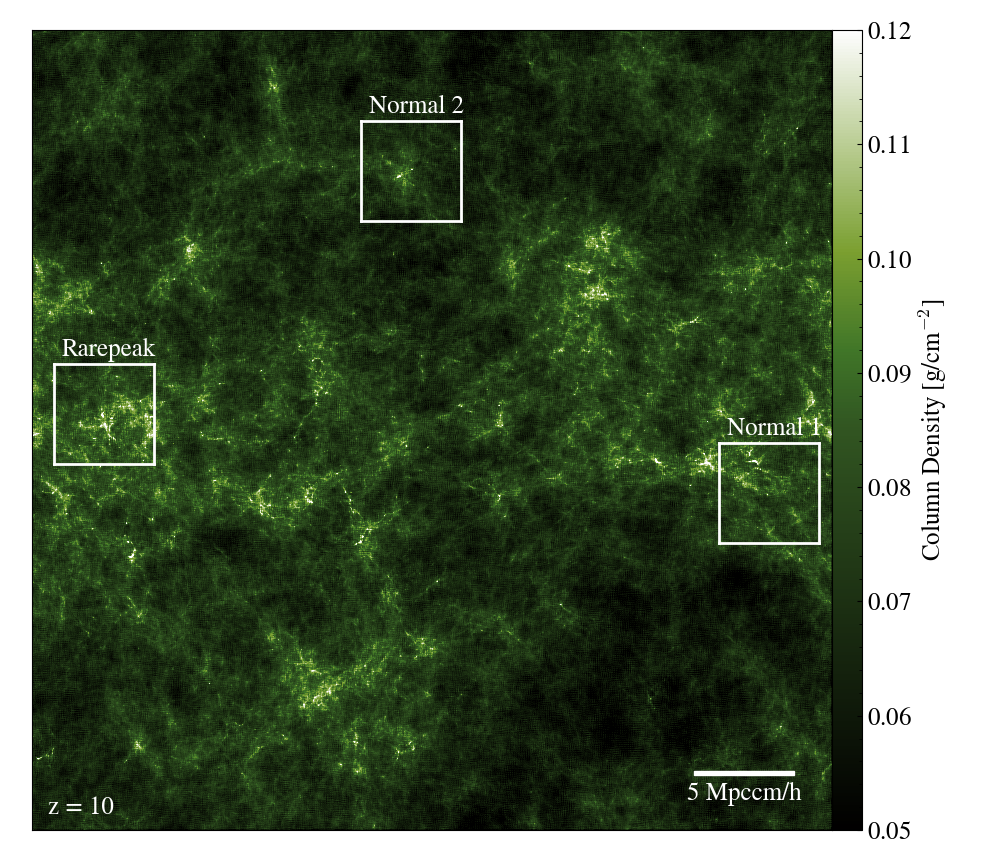}}
 
\caption[]
{\label{fig:DMViz}
 The Parent Box region (see Table \ref{table:BoxSize}) shown as a projection of the dark matter density field. The Parent Box is 40 comoving Mpc h$^{-1}$, run to $z = 10$ with only dark matter. From this box, three regions were selected - the \rarepeak region contains the most massive halo in the box at $z = 10$ (M$_{\rm Halo} = 1.8 \times 10^{11}$ \msolarc). Both of the normal regions (\NormalOne \& \NormalTwo) contain halos with masses between $1.5 \times 10^{10}$ \msolar and $2.5 \times 10^{10} $ \msolarc. The scale line in the bottom right shows 5 comoving Mpc h$^{-1}$. \\ \\}
\end{figure*}


\subsection{Dynamical Friction}
\noindent Dynamical friction is the drag force experienced by heavy particles moving through a background sea of dark matter and stars, causing the orbits to decay towards the gravitational potential minima \citep{Chandrasekhar1943, Binney2011}. In cosmological simulations, the large dark matter particle mass makes it hard to resolve gravitational friction and also leads to unphysical gravitational heating due to two-body encounters between particles. In order to avoid this unphysical heating, gravitational interactions between particles are damped below a specified gravitational softening length \citep{Governato1994, Kazantzidis2005}. However, this has the side effect of reducing the dynamical friction acting on the heavy particles. To correct for this effect, we include an additional acceleration term acting on the black hole particles that accounts for the dynamical friction occurring due to interactions within the volume in which they are artificially suppressed by the softening, namely the volume $V_{\rm s} = 4/3 \pi r_s^3$,
where $r_s$ is the softening length. 

\indent We build our dynamical friction model based on \cite{Tremmel2015}, adding it to \textsc{arepo2}. Calculating dynamical friction in \arepo\  is non-trivial because not all particles are active on every timestep. We therefore implement a scheme where dark matter particles in the vicinity of each MBH are identified on the global timestep when all particles are active, and assign them to the same time bin as the MBH particle. This ensures that the local dark matter distribution surrounding each MBH is evolved on the same timestep, allowing a well-defined estimate of the background density entering the dynamical friction calculation. We assume that the velocity dispersion of particles within the radius $r_s$ is isotropic and that the classical Chandrasekhar treatment of dynamical friction applies \citep{Chandrasekhar1943}. We further assume that only particles moving slower than the MBH would influence its motion, giving us an acceleration of the form
\begin{equation}
    a_{\rm DF} = - 4 \pi G^2 M_{\rm BH} \rho(<v_{\rm BH})\ln{\Lambda} \frac{v_{\rm BH}}{|v_{\rm BH}^3|}
\end{equation}
where $\rho(<v_{\rm BH})$ is the density of background particles moving slower than than the MBH, estimated from the total dark matter mass within the smoothing volume, and $v_{\rm BH}$ is the velocity of the MBH. The Coulomb logarithm, $\ln{\Lambda}$,  depends on the maximum and minimum impact parameters, $b_{\rm max}$ and $b_{\rm min}$ such that $\Lambda = \frac{b_{\rm max}}{b_{\rm min}}$. Since gravitational interactions beyond the softening length are well resolved, we take $b_{\rm max} = r_{\rm s}$ in order to avoid any over-counting of the effects of dynamical friction. For the minimum impact parameter, we take the radius of the influence for the MBH, i.e.
\begin{equation}
    b_{\rm min} =  \frac{GM_{\rm BH}}{v_{\rm BH}^2}.
\end{equation}
For the purpose of this study, only dynamical friction from dark matter particles is taken into consideration when computing this correction. The dynamical friction from stars, will be small at our current resolution and was neglected for this work but will be added in future versions of the model. In addition, we only employ dynamical friction for MBHs whose mass is larger than 5 times the dark matter particle mass ( $\sim 3.45 \times 10^5 h^{-1}$ \msolarc), as the dynamics can be hindered further by applying the correction when M$_{\rm BH} \sim $M$_{\rm DM}$ \citep{Tremmel2015}. For MBHs with masses comparable to individual dark matter particles, the resulting dynamical friction becomes noisy and dominated by stochastic fluctuations \citep{Tremmel2015}. The resulting acceleration is added to the MBH acceleration to be integrated over the next time-step. As the mass of the MBH increases, the radius of influence grows and can eventually become larger than $b_{\rm max}$. Once this happens, we assume that dynamical friction is fully resolved and no longer include this correction term.

\subsection{Stellar and Black Hole Mergers}
\label{sec:mergers}
\noindent We allow \smartstar particles to merge based on the treatment originally implemented in \cite{Prole2022}. \smartstar particles are merged if:
   \begin{itemize}
      \item They lie within each other’s accretion radius of 0.62 ckpc h$^{-1}$. 
      \item They are moving towards each other.
      \item Their relative accelerations are $<$0.
      \item They are gravitationally bound to each other.
   \end{itemize}
As \smartstar particles carry no thermal data, the last criterion simply requires that their gravitational potential exceeds the kinetic energy of the system. When these criteria are met, the larger of the particles gains the mass and linear momentum of the smaller particle, and its position is shifted to the centre of mass of the system. Comparing the accretion radius of 0.62 ckpc h$^{-1}$ to the sink particle gravitational softening length of 3.1 ckpc h$^{-1}$ means that mergers will likely be somewhat suppressed within this initial suite of `low resolution' runs.

Some important features regarding the \smartstar particle types following a merger are as follows:
   \begin{itemize}
      \item Regardless of which particle has the greater mass, if one of the particles is a BH, the surviving particle will be a BH. 
      
      \item When 2 PopIII particles merge, the outcome is a PopIII particle and the stellar lifetime is reset to that of the new combined stellar mass.
      
      \item We do not allow PopII stellar cluster particles to merge with any \smartstar particle type.
      
   \end{itemize}

\begin{table*}[t]

\centering
\begin{tabular}{| P{1.7cm} | P{3cm} | P{2.0cm} | P{2.5cm} | P{2.5cm} | P{1.6cm} | P{2.0cm} |}
\hline
Name & Box size or zoom region size [cMpc h$^{-1}$] & Particle Mass [\msolarc h$^{-1}$] & Dark Matter Softening Length  [ckpc h$^{-1}$] & Gas Softening Length  [ckpc h$^{-1}$] & Minimum Cell Size [ckpc h$^{-1}$] \\
\hline
\textsc{Parent} & 40 & $4.16 \times 10^7$ & 80 & - & -\\
\hline
\rarepeak & 6 & $6.9 \times 10^4$ & 3.1  & 0.15 & 0.062 \\
\hline
\NormalOne & 6 & $6.9 \times 10^4$  & 3.1  & 0.15 & 0.062 \\
\hline
\NormalTwo & 6 & $6.9 \times 10^4$  & 3.1  & 0.15 & 0.062 \\
\hline
\end{tabular}
\caption{\label{table:BoxSize}
Runtime parameters of our simulations in co-moving units. 
}
\end{table*}
\section{Initial Conditions and Parent Box Size}  \label{sec:box} 

\begin{figure}[!t]
\centering
  \includegraphics[width=0.955\linewidth]{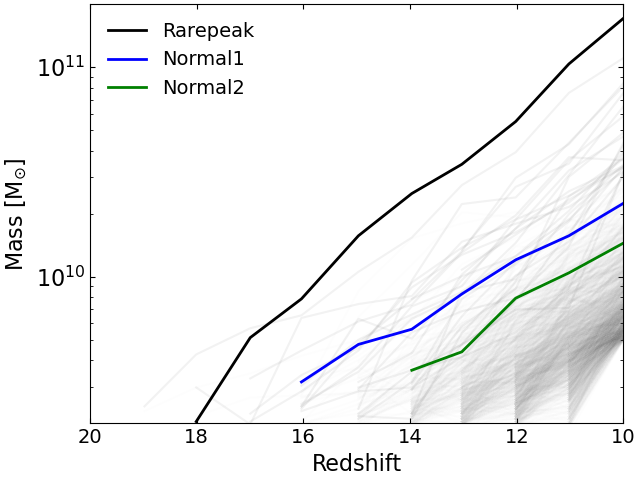}
 
\caption[]
{\label{fig:AllHaloes}
 The growth of the most massive halo in each of the target regions selected. The most massive halos are coloured and identified in the legend. Other halos from the Parent box simulation are shown in greyscale colour. Outputs from these dark matter only simulations were set at $\delta z = 1.0$ and hence there is some discrete noise effects in halo growth trajectories due to this cadence. \\}
\end{figure}

Following on from the subgrid descriptions we now describe in detail the simulation realisation, initial conditions and runtime parameters. The parent simulation box is initialised at $z=127$ with a  side length of 40 cMpc h$^{-1}$ in comoving units, generated using MUSIC \citep{Hahn2011} with cosmological parameters $\Omega_0=0.315$, $\Omega_\Lambda=0.685$, $\Omega_B=0.0486$ and $h=0.674$ \citep{planck-collaboration2020}. The parent simulation contained only dark matter and was designed to identify regions of interest for a follow-on zoom simulation setup. The details of the parent simulation box are given in Table \ref{table:BoxSize}. After running this dark matter only simulation to redshift of $z = 10$, we identified three regions for re-simulation with a full hydrodynamics treatment. These regions are shown in Figure \ref{fig:DMViz} and are denoted as \rarepeak, \NormalOne and \NormalTwo following the conventions used in the \textsc{Renaissance} suite (upon which these simulations are loosely based). 
These three regions were identified based on the relative rarity of the structures in that region. The \rarepeak region is centred on the most massive halo in the Parent box at $z = 10$. The \NormalOne and \NormalTwo regions were selected because they contain halos of  approximately $10^{10}$~\msolar at $z = 10$, which is the approximate expected mass of the most massive halo in a region of mean cosmic density at $z = 10$. For comparison, the most massive halo in the \rarepeak region has a mass of over $10^{11}$~\msolar at the same redshift. The masses of all of the halos found in the parent box are plotted in Figure \ref{fig:AllHaloes}. The most massive halo in the \rarepeak region is coloured black, the most massive halo in the \NormalOne region is coloured blue and the most massive halo in the \NormalTwo region is coloured green. Other halos in the full parent box volume are shown in greyscale.

\indent For re-simulation, the centres of each of the regions are used along with the MUSIC \citep{Hahn2011} initial conditions generator to re-generate initial conditions, each of which is centred on the regions of interest (\rarepeak, \NormalOne, \NormalTwo). Three additional levels of refinement are added resulting in a new, high resolution dark matter particle mass of $6.9 \times 10^4$ \msolarc h$^{-1}$. Additionally, the spatial resolution of the grid is increased, the minimum cell size is set to 62 comoving pc h$^{-1}$ ($\sim$ 6 pc at $z=15$) and the minimum softening length of the dark matter and gas are set to 3.1 comoving kpc h$^{-1}$ ($\sim 300$ pc physical at $z = 15$) and 0.15 comoving kpc h$^{-1}$ ($\sim 14$ pc physical at $z = 15$) respectively. At this resolution, we are able to identify halos with masses a few times $10^6$ \msolar and we can spatially resolve the inner parts of galaxies down to scales of approximately 0.2 comoving kpc h$^{-1}$ (a few tens of parsecs at $z = 15$).  
\indent 

\begin{figure}
  \includegraphics[width=0.97\linewidth]{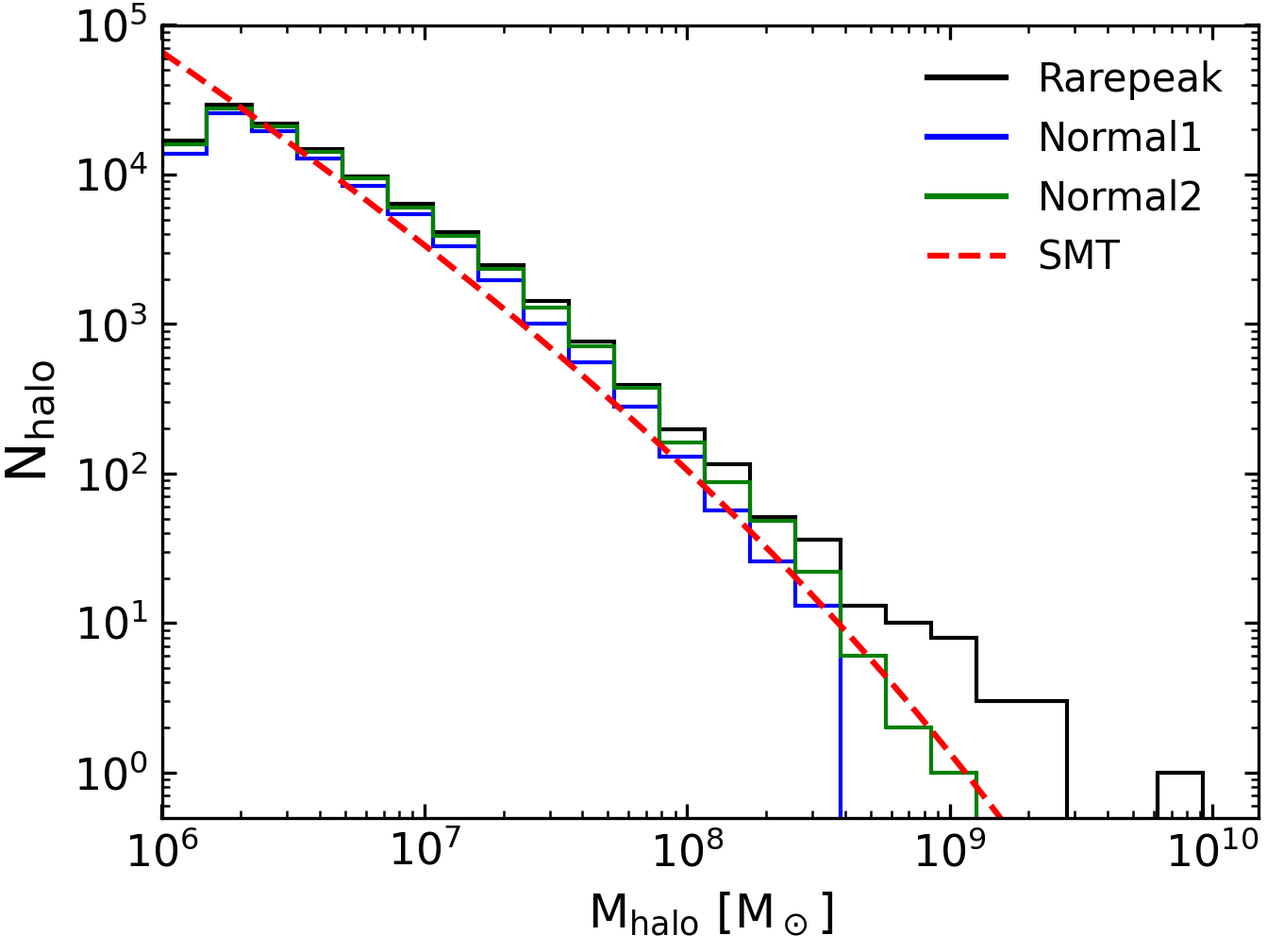}
 
\caption[]
{\label{fig:HMF}
 Halo mass functions for the \rarepeak, \NormalOne and \NormalTwo regions at $z=15$. The halos were identified using the Subfind halo finder. Overlaid (dashed red) is the predicted halo mass function at $z=15$ from  \citet{Sheth2001} -- labelled ``SMT''. As expected the \rarepeak region shows a deviation from the analytical prediction at $z = 15$ at the higher mass end, reflective of the overdense environment found in that region. The \NormalOne and \NormalTwo regions show closer agreement to the expected halo distribution. \\}
\end{figure}


\begin{figure*}[!th]
\centering
  \includegraphics[width=0.8\linewidth]{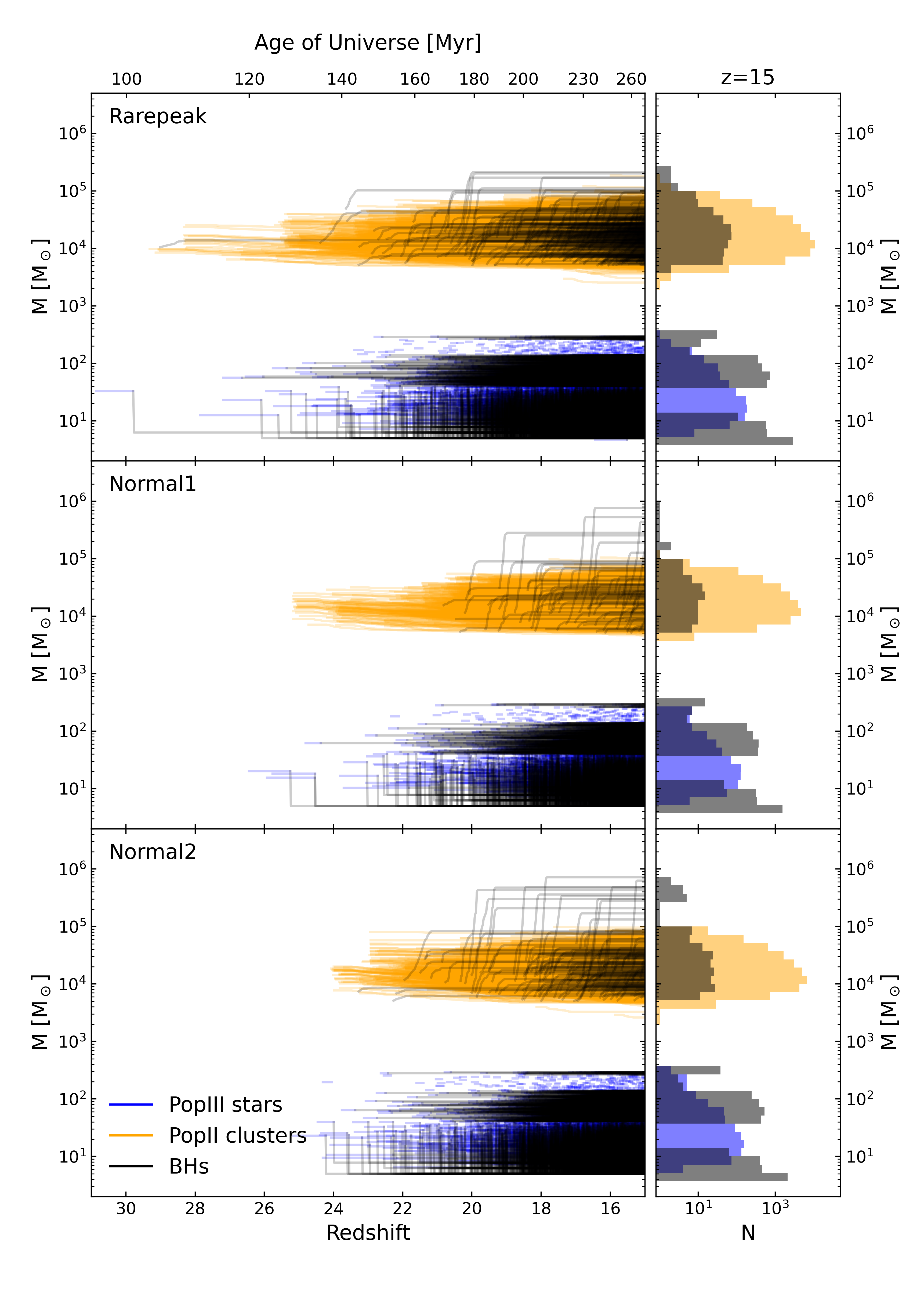}
\caption[]
{\label{fig:growth} Redshift evolution of \smartstar particle masses for the \rarepeak (top) and \NormalOne (middle)  and \NormalTwo (bottom) regions, showing PopIII stars (blue), PopII stellar cluster particles (orange) and BHs (black). The left panel of each figure shows the redshift evolution while the right hand panel shows the resulting demographics at $z = 15$. The number of PopII stars is approximately the same in each region, while the number of heavy seed black holes is largest in the \rarepeak region, as expected. \\ }
\end{figure*}

\section{Results}
\label{sec:results}

\noindent The goal of this paper is to introduce the \texttt{SEEDZ} simulations suite, present the implemented numerical methods and share our first results. The results shown here are from the calibration set of simulations known as the `low resolution' suite. Two additional suites of simulations with progressively higher resolution are planned in the near future, with the  mass resolution increased by a factor of eight, and in addition the spatial resolution increasing by a factor of four in each iteration. Despite the naming convention, the mass/spatial resolution  and underlying subgrid modelling in the simulations presented here represent some of the most sophisticated simulations of the high-z Universe to date (see \S \ref{sec:discussion}, where we compare this simulation suite to others in the literature). The following results represent the simulations down to $z=15$. This set of `low resolution' simulations are still ongoing and further analysis of this suite as it evolves to its final redshift of $z=10$ will be presented in a future publication.

\begin{figure*}
  \centering
  \includegraphics[width=0.99\linewidth]{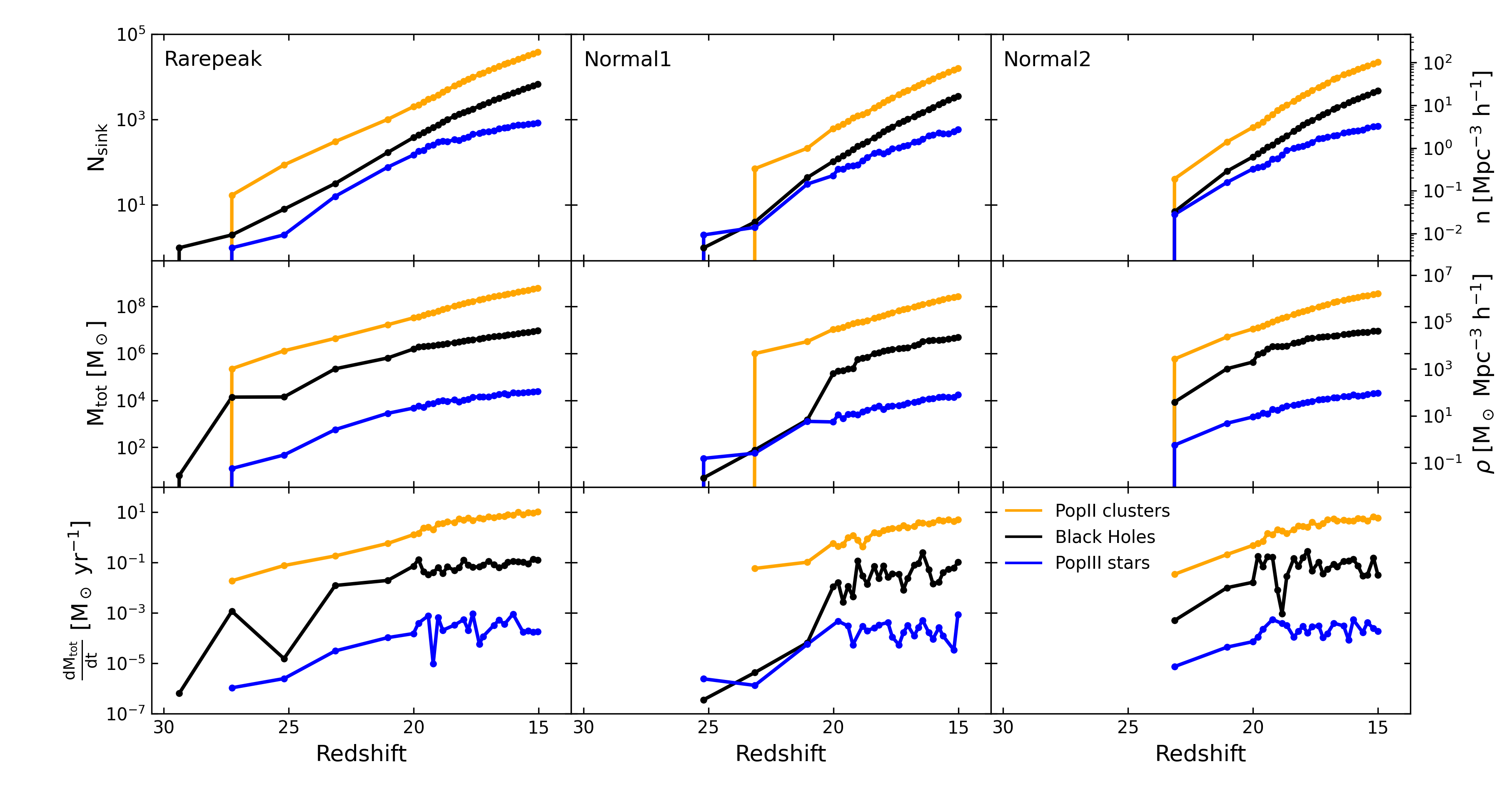}
\caption[]
{\label{fig:population} For each of the simulation boxes, we show the total mass in PopIII stars (blue), BHs (black) and PopII clusters (orange) as a function of redshift. While PopIII stars form first in all cases (see Figure \ref{fig:growth}), PopII star formation quickly dominates over the PopIII population. The mass in BHs is approximately 2 orders of magnitude below the mass in stars, with the mass growth in PopII stars exceeding that in BH by approximately 2 orders of magnitude as well. In the top 2 rows, we provide a secondary $y$-axis, showing the number density and mass density for these objects. \\ }
\end{figure*}

\begin{figure}
  \centering
  \includegraphics[width=\linewidth]{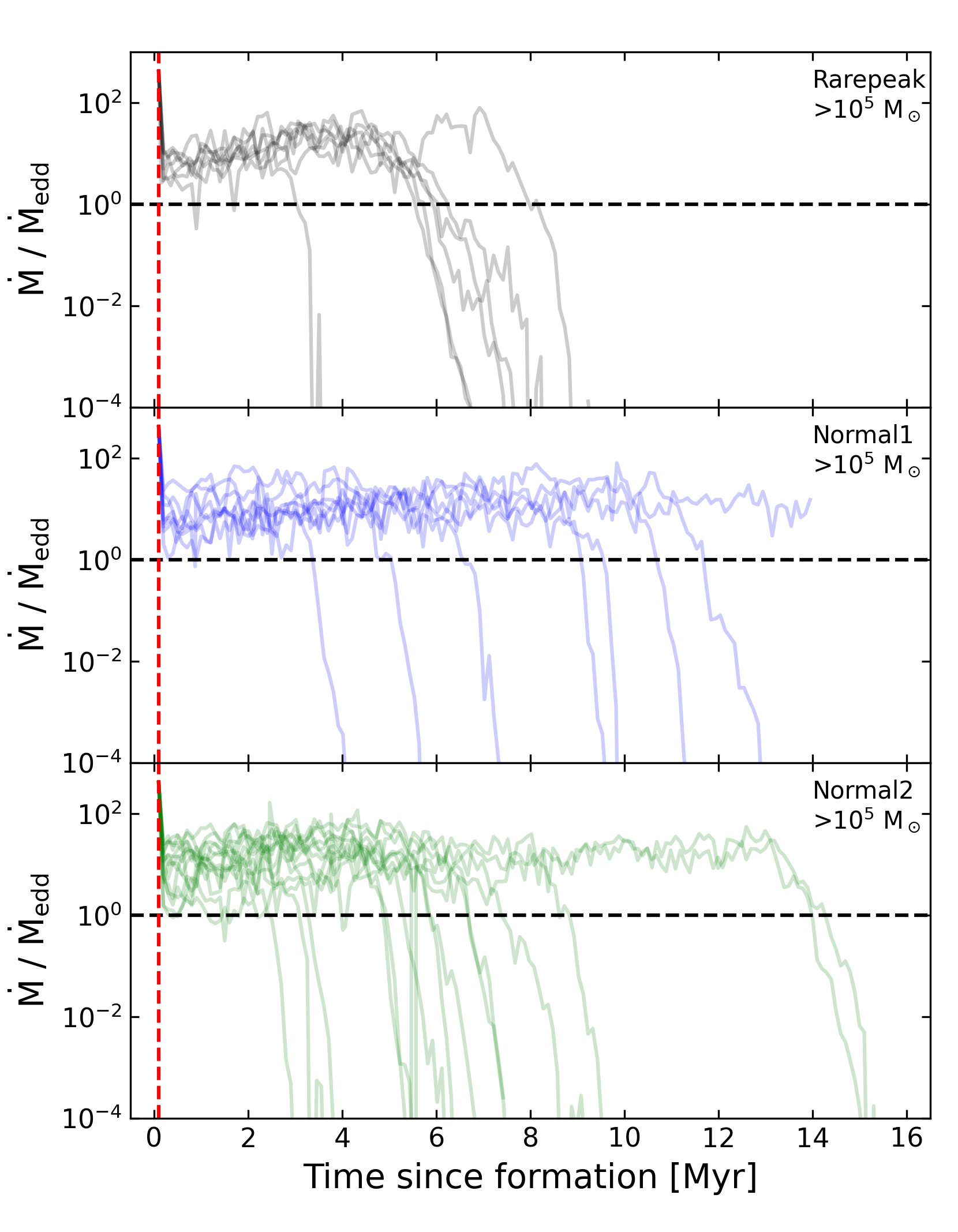}
\caption[]
{\label{fig:eddington} The growth of all heavy seeds with final masses above 10$^5$ M$_\odot$ as a function of time since their formation, expressed as the ratio of their accretion rate to the Eddington accretion rate, for the \rarepeak, \NormalOne and \NormalTwo regions. The black dashed line shows Eddington rate accretion, while the red dashed line shows the point where thermal accretion feedback kicks in.\\ }
\end{figure}

\begin{figure}
  \centering
  \includegraphics[width=\linewidth]{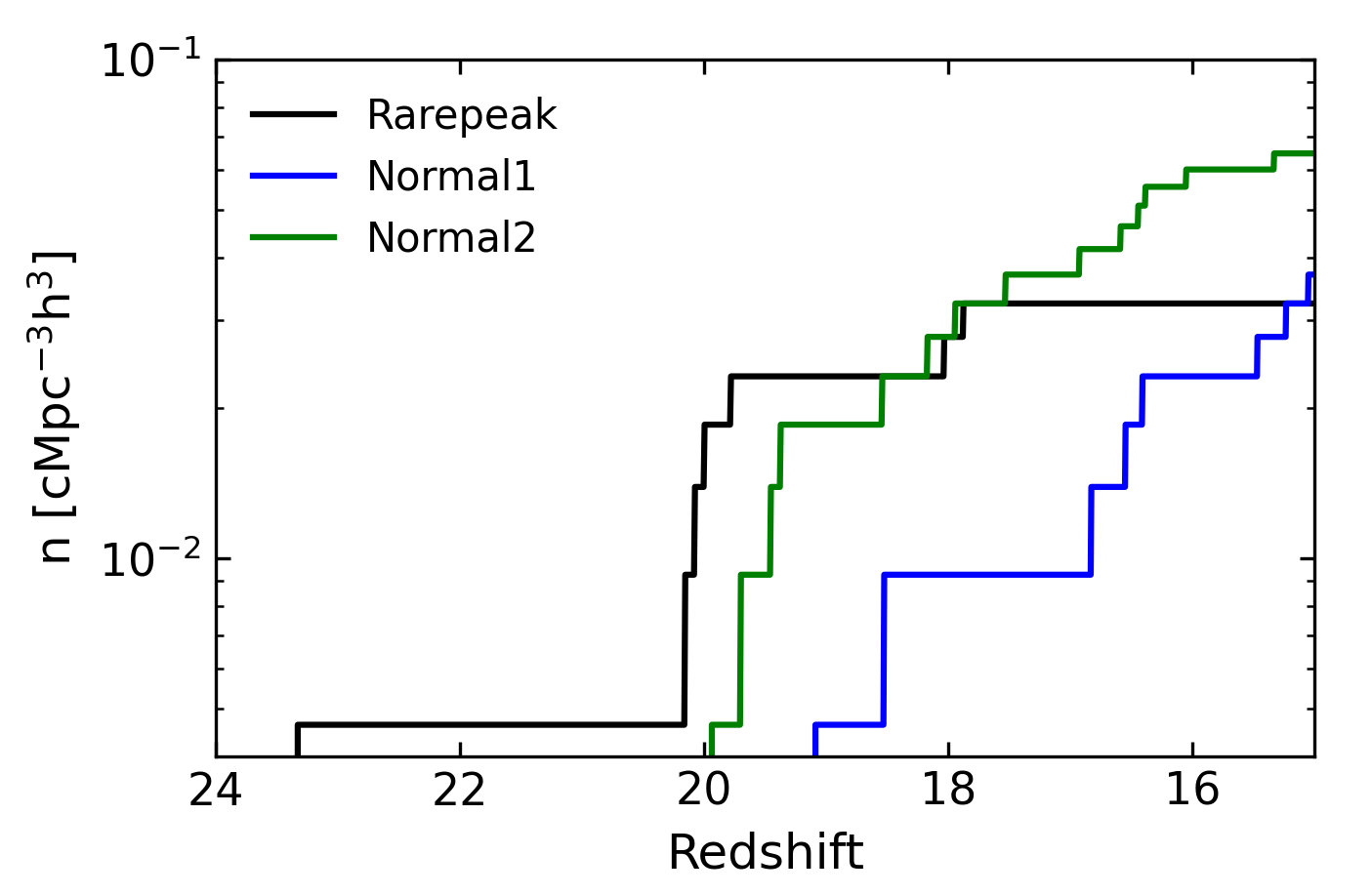}
\caption[]
{\label{fig:numberdensity} Number density of black holes with masses greater than 10$^5$ M$_\odot$ as a function of redshift, for the \rarepeak, \NormalOne and \NormalTwo regions. The zoom regions of the simulation have side lengths of 6 cMpc h$^{-1}$, giving a volume of 216 cMpc$^{3}$ h$^{-3}$. By $z=15$, these regions produced 7, 8 and 14 BHs above 10$^5$ M$_\odot$, respectively.\\ }
\end{figure}

\subsection{Global Simulation Results}

\noindent To validate the underlying physics of gravitational collapse in our simulations, we investigate the global structure formation metrics in Figure \ref{fig:HMF}, which shows the halo mass function (HMF) at $z=15$. These halos were identified by Subfind \citep{Springel2021}, the inbuilt halo finder module in \arepo. Halos were identified if they contained at least 20 particles. We find excellent agreement with the predicted HMFs from \cite{Sheth2001}. The \rarepeak simulation (black line) shows an obvious deviation from the theoretical curve at high halo masses, as this \rarepeak zoom is centred on the most massive halo found at $z = 10$ in a much larger volume. Both the \NormalOne and \NormalTwo  zoom simulations agree somewhat better with the theoretical fit as expected. Note that while there is a deviation also at the lower mass end, this is simply due to our limited resolution of halos below a mass scale of approximately $10^6$ \msolar  (our DM particle resolution is $6.9 \times 10^4$ \msolarc h$^{-1}$)

The new subgrid models introduced into \arepo\  revolve around our \smartstar prescription \citep[see also][]{Regan2018a, Regan2019, Regan2020}. Figure \ref{fig:growth} shows the formation and evolution of the different \smartstar particle types across cosmic time. While the formation of the first PopIII stars (blue) begins at $z\sim25$ in the normal regions, aligning with many other studies \citep[e.g.][]{Chen2014, Xu2016a}, the overdensity represented by our Rarepeak region collapses earlier, allowing the first stars to form from $z\sim31$. In all regions, the occurrence of the first SN explosion and subsequent metal enrichment is quickly followed by the formation of the first PopII cluster particles (orange) \citep{Smith2007}. Our PopII star clusters form with masses in a range between a few times $10^3$ \msolar and approximately $10^5$ \msolarc. We do not prescribe these masses, they are instead derived from the mass resolution of cells identified as star forming. At higher resolution, the masses of the star cluster particles would inevitably be lower, ultimately reaching individual PopII stars at extreme resolutions. After the first SNe explosions, PopIII stars shed mass and light seed BHs (black) are birthed from the remnants (and hence the reduction in mass compared to the progenitor PopIII star). Note that not all PopIII stars result in a remnant (light seed) black hole (see \S \ref{sec:SNe}). While, high resolution simulations using our \smartstar model have shown that these light seeds are capable of accreting up to $10^5$ \msolar \citep{Mehta2024}, as also shown by other authors using different subgrid models \citep{Shi2023, Shi2024, Shi2024a, Gordon2025, Zana2025}, this set of calibration simulations lacks the necessary resolution to produce this growth (see also Mehta et al., in prep). As a result, the light seed BHs remain at their birth mass and do not grow in our simulations. However, the resolution is sufficient to resolve accretion onto our heavy seed BHs, which form with initial masses in the range $3 \times 10^3-10^5$ M$_\odot$ (see \S \ref{Sec:seeding}).

\indent Heavy seeds form in our simulation suite beginning at $z \sim 25$ in our \rarepeak region and $z \sim 20 - 24$ in the \NormalOne and \NormalTwo regions. In all cases growth is typically vigorous initially (within the first few tens of Myr) before growth stalls. In a companion paper \citep{Prole2026}, we show that the sharp cut-off in accretion onto these BHs is caused by accretion feedback from the BH itself, rather than exhausting the accretion supply or from nearby SNe explosions. However, a detailed examination of all of their properties is outside of the scope of this introductory paper. 

While the first heavy seeds form earlier in the \rarepeak region, our most massive BHs at $z=15$ reside within the {\sc Normal} regions despite all simulations having the same formation, evolution and feedback implementations. In the right hand column of Figure \ref{fig:growth}, we show the total number of black holes, PopII clusters and PopIII stars at $z = 15$ (see Figure \ref{fig:population} for an evolution of the number of objects against redshift). The total number of heavy seed black holes at $z = 15$ is 643 (total mass in heavy seed black holes in $\sim2\times 10^7$ \msolarc), the total number of PopII clusters is 75729 (total mass in PopII stars is $\sim10^9$ \msolarc), the total number of PopIII stars is 2126 (total mass in PopIII stars is $\sim6\times 10^4$ \msolarc) and finally the total number of light seed black holes is 14367 (total mass in light seed black holes is $\sim4\times 10^5$ \msolarc).

The top two panels of Figure \ref{fig:population} show how the total number and mass in each of the \texttt{smartstar} populations evolves over time. Note that while the first objects to form in each of the simulations were PopIII stars, due to the time separation between snapshots this is not reflected in the graph, which first picks up these objects after they have transitioned into BHs. Following the formation of the first PopIII stars and BHs, the Universe becomes dominated (in mass) by metal-enriched PopII clusters. The bottom panel of Figure \ref{fig:population} shows the growth rate (in \msolarc/yr$^{-1}$) of each species. The formation rate of PopIII stars begins to flatten as the simulation box becomes more metal enriched. PopIII formation is expected to end completely by $z\sim4$ \citep{Wise2011,Johnson2013a,Xu2016a,Sarmento2018}.
\\ 
\subsection{Massive Black Hole Growth}

\noindent In Figure \ref{fig:eddington}, we show the growth rate of all heavy seed BHs with final masses (at $z = 15$) above 10$^5$ M$_\odot$, expressed as the ratio of their accretion rate to their Eddington accretion rate. Despite the onset of thermal feedback 10$^5$ yr after their creation, we find that super-Eddington accretion can occur for up to $\sim$1-15 Myr after formation. Sustained super-Eddington growth has been produced in similar simulations for the order of 100 kyr \citep{Gordon2025} up to tens of Myr \citep{Lupi2024}, and is supported by a growing number of observations \citep{Du2018,Yue2023,Bhatt2024,Jin2024,Suh2025,Yang2024,Lambrides2024}.

Figure \ref{fig:numberdensity} shows how the number density of MBHs (with a final masses greater than $10^5$ M$_\odot$) evolves with redshift. Each of the simulated regions produces a number density of a few times $\sim$10$^{-2}$ cMpc$^{-3}$ h$^3$ by $z=15$, with the \NormalTwo region achieving roughly a factor of two higher than the other regions. This corresponds to 7, 8 and 14 MBHs in the \rarepeak, \NormalOne and \NormalTwo regions, respectively. The  factor of $\sim$2 difference between the normal regions is likely due to cosmic variance. These numbers are roughly consistent with previous numerical estimates \citep{Trinca2022,Chiaki2023,McCaffrey2024}, although they are significantly higher than LW channel models \citep{OBrennan2025}. 

To compare these numbers with observations, the SMBH mass spectrum appears to be a decreasing function with BH mass. JWST/NIRCam fields show $z=4-9$ number densities of around 10$^{-4}$ Mpc$^{-3}$ for 10$^6$ M$_\odot$ BHs, decreasing to 10$^{-6}$ Mpc$^{-3}$ for 10$^8$ M$_\odot$ BHs \citep{Kokorev2024}, consistent with the $z=4$ quasar mass function \citep{He2024}. The overall number density of LRDs is estimated to be 10$^{-4}$ Mpc$^{-3}$ \citep{Perez-Gonzalez2024}, meaning only $\sim0.2\%$ of our $>10^5$ M$_\odot$ BHs would need to continue growing to be consistent with observations.

\subsection{Halo and Galaxy Properties}

\begin{figure}
  \includegraphics[width=\linewidth]{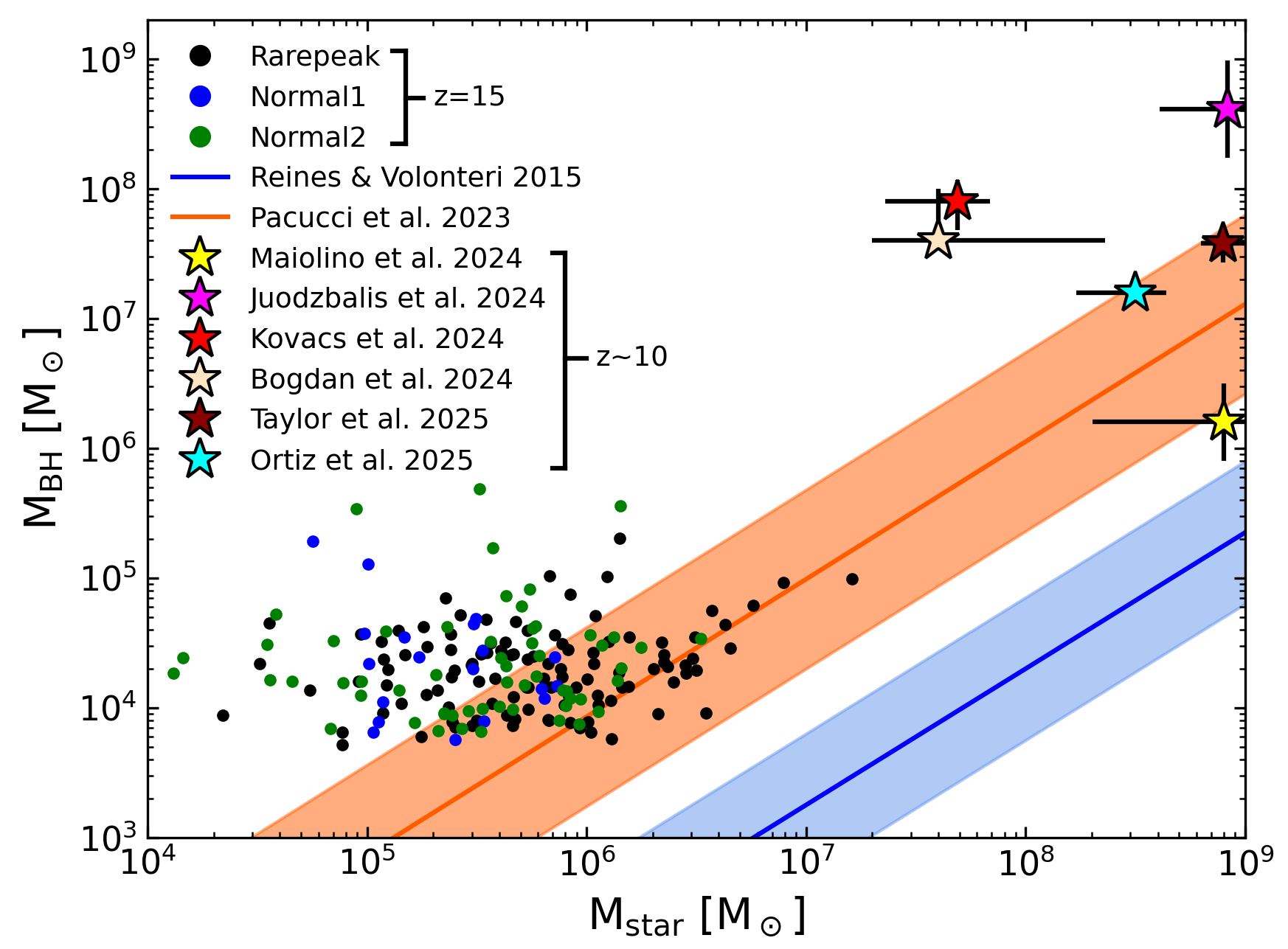}
\caption[]
{\label{fig:bh_ratios} Relations between the largest BH mass per halo and the total stellar mass within the halo at $z=15$. Also shown are the observationally derived local galaxy ($z$=0) relation from \cite{Reines2015} and the high redshift ($z$=4-7) relation from \cite{Pacucci2023}. We also include $z\sim 10$ BH observations from \cite{Maiolino2024} , \cite{Juodzbalis2024}, \cite{Kovacs2024}, \cite{Bogdan2024}, \cite{Taylor2025} and \cite{Ortiz2025}.\\}
\end{figure}

\begin{figure}
  \centering\includegraphics[width=\linewidth]{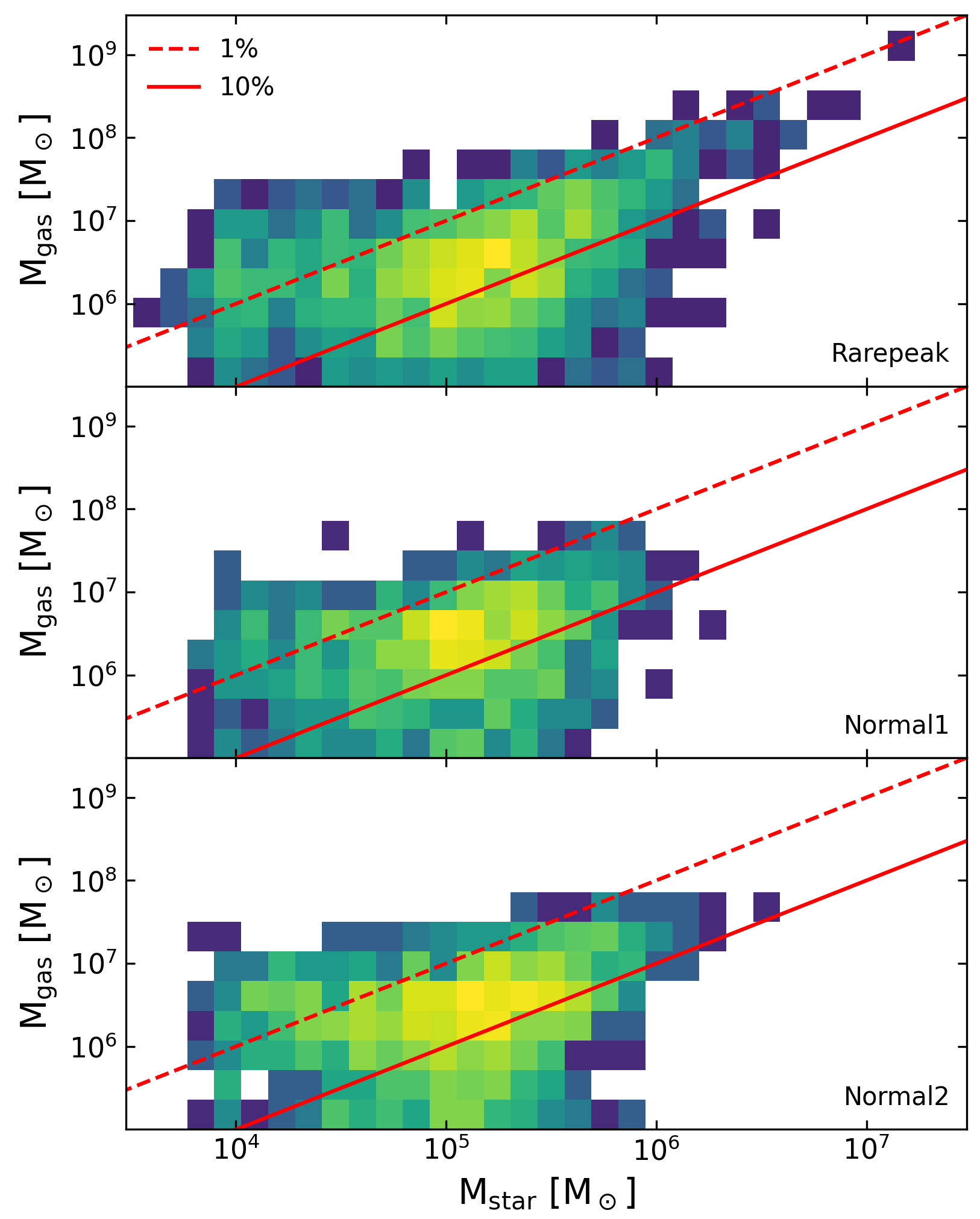}
\caption[]
{\label{fig:gas_ratios} Relation between halo gas mass and stellar mass i.e. star formation efficiency (SFE). We present this as a 2-dimensional heat map, where dark blue and bright yellow colours represent low and high numbers of galaxies, respectively. We plot lines for SFEs of 1$\%$ (dashed) and 10$\%$ (solid). \\}
\end{figure}

\begin{figure*}
\centering
  \includegraphics[width=0.9\linewidth]{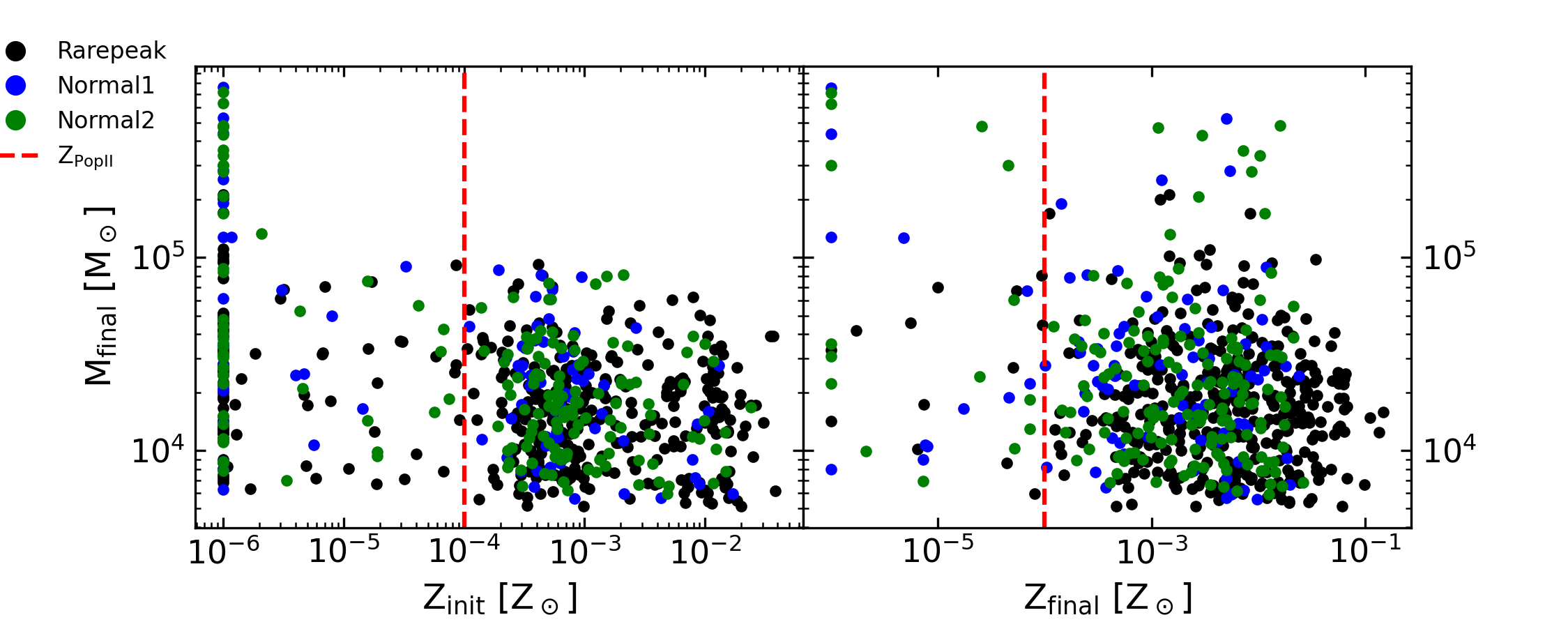}
\caption[]
{\label{fig:metals} For all heavy seed BHs ($>1000$ M$_\odot$), we show the initial metallicities at the time of their formation versus their mass at $z=15$. The red dashed line shows the threshold metallicities for PopII cluster formation. Most of our heavy seed black holes form with initial metallicities between $10^{-4}\ \rm{Z}_{\odot}$ and $10^{-2}\  \rm{Z}_{\odot}$ and live in environments of metallicities between $10^{-4}\  \rm{Z}_{\odot}$ and $10^{-1}\  \rm{Z}_{\odot}$ at $z = 15$.}
\end{figure*}

\noindent In Figure \ref{fig:bh_ratios} we investigate the relation between a galaxy's most massive BH and its stellar mass, using the subfind halo positions and radii to approximate the galaxy. The radius of the galaxy is taken to be the radius containing half the mass of the halo. In orange we plot the recently derived BH - stellar mass relationship found for high-redshift BHs found with JWST \citep{Pacucci2023}. In blue we show the local ($z = 0$) scaling relations (Equation 4 from \citealt{Reines2015}). What is clear is that our galaxies and BHs do not show a well defined trend and are not well described by either the \cite{Pacucci2023} or the \cite{Reines2015} relationships when extrapolated down to the mass regime relevant to our current results. This is not surprising. The BHs in our simulations form with heavy seed masses and in some cases accrete very rapidly - leading in many cases to MBH masses in excess of the host stellar masses. Referencing back to Figure \ref{fig:population} we see that the PopII masses are growing at rates significantly in excess of the MBH masses, and we therefore predict that our galaxies will asymptote to the \cite{Pacucci2023} relationship, and eventually (as mature galaxies) to the \cite{Reines2015} relationship.  We compare our simulated galaxies at $z=15$ with a number of the highest redshift ($z\sim10$) MBH observations (shown as star markers) that have stellar mass estimates. We expect that our MBHs and host galaxies to move onto the high redshift relation to match these observations as the simulations reach $z=10$.

Figure \ref{fig:gas_ratios} shows the relation between the gas mass in halos and their total stellar mass, i.e.\ the star formation efficiency (SFE) in these early galaxies. We present this as a 2-dimensional heat map, where blue and yellow colours represent low and high numbers of galaxies, respectively. Towards higher halo masses ($M_{\rm star} >10^6$ M$_\odot$), these galaxies have SFEs of between 1-10$\%$, which is consistent with observed \citep{Krumholz2007,Sun2016, Barnes2017} and numerical estimates \citep{Behroozi2012,Grudic2022}. It is important to note that SFE is not typically a fixed value, but an increasing function with halo mass. For example, the \textsc{Renaissance} simulations \citep{Chen2014,Xu2016,Xu2016, Wise2019,oshea2015} typically produce SFEs of 0.1$\%$ in 10$^7$ M$_\odot$ galaxies, increasing to $\sim5\%$ for 10$^9$ M$_\odot$ galaxies, with extreme cases exceeding SFE$>10\%$ \citep{McCaffrey2023}. 

In Figure \ref{fig:metals}, we show our heavy seed BH masses at $z=15$, plotted against the initial metallicity of their host gas cell at the time of their formation (left). We contrast this to the metallicity surrounding these objects (within 20~pc) at $z=15$ (right). While most of these heavy seeds form in relatively high metallicity environments (${\rm Z} > 10^{-4}$ Z$_\odot$), we find that all of the BHs that grew above $10^5$ M$_\odot$ actually formed in low metallicity  ($\ll 10^{-4}$ Z$_\odot$) environments. the RHS panel shows that while the metallicity surrounding most of these growing BHs increased throughout their lives, a few of the most massive BHs remained embedded within low metallicity environments. These BHs have parallels with recent observations of SMBHs in near-pristine galaxies \citep{Juodzbalis2024, Maiolino2025}.

\indent Our formation criteria for heavy seed black holes is agnostic to the metallicity criteria - we form heavy seed MBHs in regions undergoing rapid gravitational collapse irrespective of the metallicity content. Our methodology accounts for the formation of heavy seeds via the formation pathways of super-massive stars and the collapse into a heavy seed black holes within a dense stellar environment. Both of these scenarios will continue to operate in metal-poor regions (Z $\gtrsim 10^{-2} Z_{\odot}$)  as defined here and perhaps also at higher metallicities. If our simulations ultimately find a large over-abundance of MBHs, beyond that observed, this may tell us that MBH formation is disfavoured in high metallicity environments. However, we do not observe that trend as of yet. We will return to this point in \S \ref{sec:discussion}.

\subsection{The Most Massive Halo}
\begin{figure*}
  \includegraphics[width=1.05\linewidth]{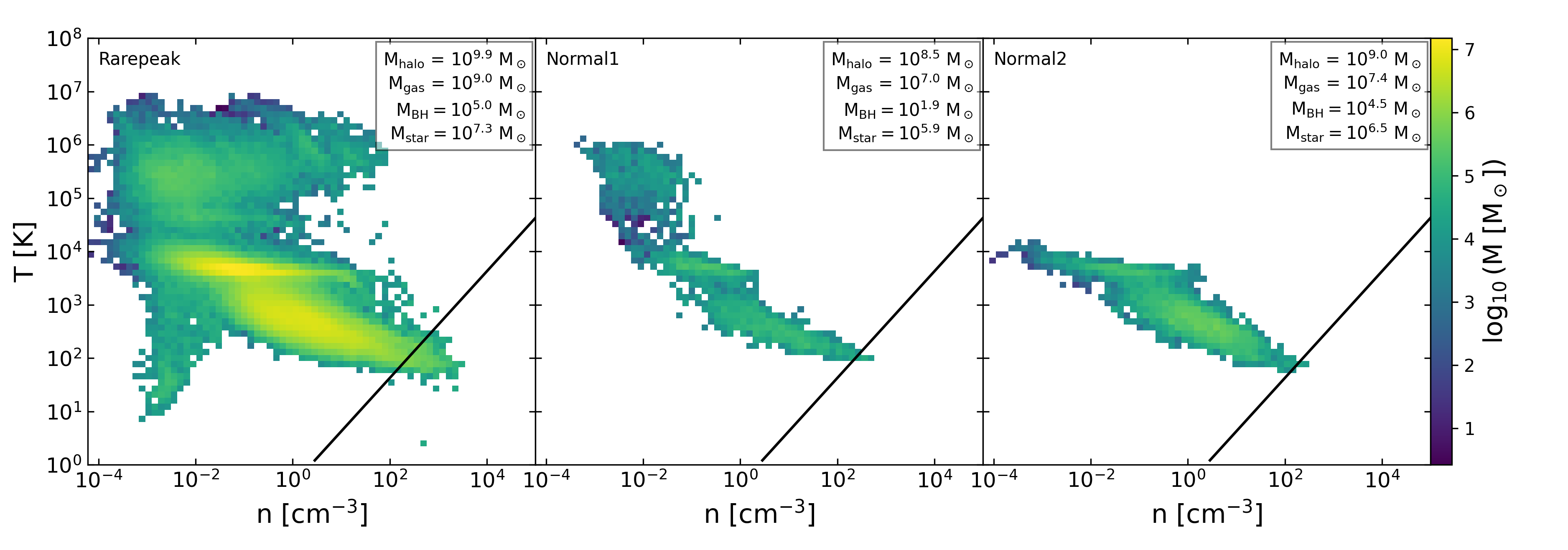}
\caption[]
{\label{fig:phase} Density versus temperature phase diagrams for the most massive halo at $z=15$ for the \rarepeak (left), \NormalOne (middle) and \NormalTwo (right) regions. The solid black line shows the limit of gravitational instability at the minimum cell length of 6 physical pc at $z=15$. We include values for the total halo mass, gas mass, largest BH mass and total stellar mass.}
\end{figure*}

Here we investigate the properties of the most massive halo in each of the simulations. Figure \ref{fig:phase} shows phase diagrams of temperature versus number density within the virial radius, along with values for the total halo mass, gas mass, largest BH mass and total stellar mass. The \rarepeak region produces a significantly more massive halo than the {\sc Normal} regions (by $z = 15$), with a total mass $\sim 10^{10}$ \msolar with a gas component of approximately 10$\%$, in line with the cosmic baryon to dark matter mass ratio. The most massive BH is  $10^5$ M$_\odot$ and the halo has a stellar mass of $\sim 2 \times 10^7$ M$_\odot$, giving a SFE of approximately 1$\%$. 

The temperature-density relationship shows three distinct phases above number densities of $10^{-2}$ cm$^{-3}$: the cold gas tail which decreases in temperature with increasing density due to H$_2$ cooling, a hotter isothermal phase at $\sim 10^4$~K typical of gas cooling via atomic Lyman-$\alpha$ emission, and a very hot phase at $10^5-10^6$ K, which is caused by a combination of SN feedback from the stellar component and accretion feedback from the heavy BH seed growth.

The most massive halos in the \NormalOne and \NormalTwo regions are 25 and 10 times less massive than the most massive halo in the \rarepeak region, respectively. The largest black hole found in the most massive halo in the \NormalOne region only has a mass of a few tens of M$_\odot$, indicating that no heavy seeds have formed within this halo. On the other hand, the most massive halo in the \NormalTwo region contains a heavy seed BH with a mass of a few times $10^4$ M$_\odot$. In all cases, the most massive BH in the simulation does not reside in the most massive halo in the region, as comparison with Figure~\ref{fig:growth} makes clear.

\begin{figure*}
  
  \centering
  \includegraphics[width=\linewidth]{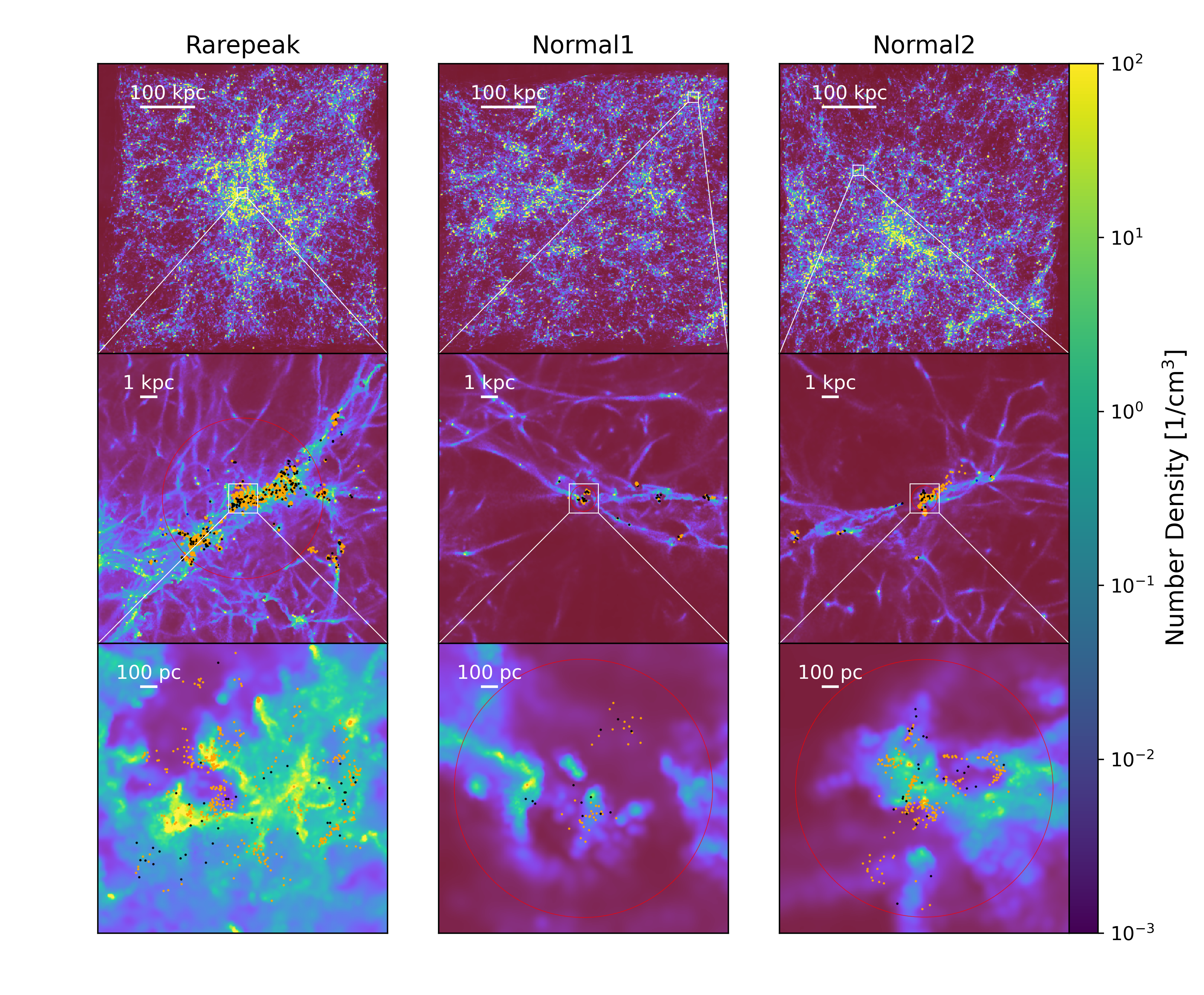}
\caption[]
{\label{fig:proj_halo} Mass weighted gas density projections of the largest halo at $z = 15$ in the \rarepeak (left), \NormalOne (middle) and \NormalTwo (right) regions. The top panel shows the entire zoom-region of the simulation ($\sim$600 physical kpc at $z=15$). Moving from top to bottom, we then show progressive zoom-ins into the most massive halo at 20 kpc and 2 kpc scales.  Aligned with the previous colour scheme, we show BHs as black dots (both light and heavy seeds), PopII cluster particles as orange dots and PopIII stars as blue dots.}
\end{figure*}

The locations of the most massive halo in each of the simulations are shown in Figure \ref{fig:proj_halo}, along with zoom-in projections around the central 20 kpc and 2 kpc. Interestingly, while the most massive halo, at $z = 15$,  resides at the center of the box in the \rarepeak region, it forms off-center in the {\sc Normal} regions, and close to the edge of the zoom-region in the \NormalOne region. This is despite the fact that the zoom-in regions were centred on the most massive halo in each region at $z = 10$. For both the \NormalOne and \NormalTwo regions, a halo of comparable mass exists in the centre of each region. In all cases, the effect of our feedback prescription is clearly visible, in the form of  pockets of under-dense gas surrounding groups of PopII cluster particles, where gas densities have been dramatically reduced following SNe explosions.

\section{Discussion}
\label{sec:discussion}
\noindent The simulations presented here introduce a new BH seeding model in cosmological simulations, which represents a marked improvement in black hole seeding and growth compared to the previous generation of similar simulations. In particular we follow the formation of light and heavy seeds. Light seeds are formed from the end point of PopIII stars, while heavy seeds form in regions that are both gravitationally unstable and experiencing mass inflow rates in excess of 1 \msolarc yr$^{-1}$. We place no dependence on metallicity or H$_2$ fraction when modelling the formation of heavy seeds - we do this to account (as generally as possible) for the variety of proposed heavy seed formation pathways (super-massive star formation, collisions within a dense stellar cluster, or the core collapse of a black hole cluster). While we also follow the formation of light seeds, our current simulation suite does not have the mass or spatial resolution to capture their growth. Hence, in this simulation suite we observe that light seeds do not grow. This does not mean that light seeds cannot grow, as their growth has been shown in higher resolution simulations \citep[][]{  Shi2023, Shi2024,Mehta2024, Shi2024a, Gordon2025, Zana2025}. 

While it is clear that the environment in which a MBH forms in will determine its seeding mechanism and growth prospects, this is often not possible to capture in large-scale cosmological simulations e.g. \texttt{Eagle} \citep{Schaye2015}, \texttt{Illustris} \citep{Vogelsberger2014, Vogelsberger2014a}, \texttt{IllustrisTNG} \citep{Pillepich2018,Springel2018}, \texttt{SIMBA} \citep{Dave2019}, \texttt{Astrid} \citep{Bird2022}, \texttt{XFable} \citep{Bigwood2025}. 
These large volume simulations have, due to computational expense, had to sacrifice complexity for computational efficiency. Hence, in all cases the MBHs in these large volume simulations have been seeded at fixed seed masses, based largely on the halo mass (with one MBH per halo) while also pinning the MBHs to the halo centre of mass (a technique known as repositioning). While advantageous in terms of reproducing local scaling relationships (e.g. $\rm{M_{BH} \  vs. \ M_*}$) this method comes at the cost of emergent behaviour (particular at high-z) as well as likely incorrect MBH merger rates and merger times \citep{Buttigieg2025}. 
More recently, higher resolution simulations have attempted to move away from these relatively simplistic treatments in both 
their treatment of the seeding criteria and their use of subgrid dynamical friction modelling, to more faithfully reproduce the MBH trajectories within host galaxies. Examples of simulations that have implemented subgrid dynamical friction modelling as well as prescriptions for heavy seed MBH formation modelling include
\texttt{Romulus} \citep{Tremmel2017}, \texttt{Marvelous} \citep{Bellovary2019}, \texttt{NewHorizon} \citep{dubois2021}.  
 The \texttt{Renaissance} simulations \citep{Chen2014,Xu2016,Xu2016, Wise2019,oshea2015} and \texttt{Serra} simulations \citep{Pallottini2022, Pallottini2023, Kohandel2025} -although ambitious in resolution- did not include MBHs. Perhaps the most advanced simulation exploring MBH seeding and growth to date is the \texttt{BlackDemon} suite \citep{Regan2023}, which self-consistently seeded MBHs based on prior PopIII stellar masses sampled from an IMF as well as due to supermassive star formation. However, the \texttt{BlackDemon} simulations were only able to simulate to $z \sim 20$ and hence were not able to sufficiently connect their results with JWST observations at $z \sim 6 - 12$ onward.

The most recent simulation suites to include sophisticated black hole seeding prescriptions include the \texttt{BRAHMA} and \texttt{MELI$\odot$RA} simulation suites. We therefore explicitly compare our simulations with these more recent works. The \texttt{BRAHMA} simulations \citep{Bhowmick2024, Bhowmick2024a, Bhowmick2025, Bhowmick2025a, Bhowmick2025b} run a range of simulation volumes and seeding criteria. For their star and black hole feedback and accretion prescriptions, they adopt those from \texttt{IllustrisTNG} as well as the \texttt{IllustrisTNG} cooling and ISM modelling. Their MBH seed masses vary from approximately $2 \times  10^3$ \msolar in their highest resolution case to $10^5$ \msolar in their larger volume, coarser resolution realisations. Again due to their more widely varying box sizes, their resolution is for some realisations somewhat better than ours (by a factor of 2-3) and at other times somewhat coarser (by over an order of magnitude). However, the major variation between their models and ours is at the subgrid level. We implement a full non-equilibrium chemical model capable of capturing both the cold and warm/hot phases in the ISM, while \texttt{BRAHMA} uses an equation of state parametrisation which misses the cold phase (e.g. \citealt{Prole2024}) but is more computationally efficient (and means they can explore a wider parameter space). The other key difference is in the BH seeding strategy adopted by both models. In \texttt{SEEDZ} we adopt a seeding strategy based on the mass inflow onto an already gravitationally unstable cell/region - capturing the strong inflows required for SMS and dense stellar cluster formation. \texttt{BRAHMA} instead implements a MBH seeding strategy based on the LW flux incident, a density threshold and a metallicity threshold.  A potential drawback of the LW flux threshold is that a large LW flux is likely unable to match recent observational constraints, and more recent \texttt{BRAHMA} results favour a highly lenient LW flux threshold \citep[see also][]{OBrennan2025}. Directly comparing our results with \texttt{BRAHMA} at this early stage of the \texttt{SEEDZ} simulation ($z = 15$) is challenging given that many of the \texttt{BRAHMA} simulations have been run to $z = 0$. However, there are a number of areas in which the simulations agree. Firstly, both simulations predict a population of over-massive black holes at $z \gtrsim 8$ with number densities (for the lenient seeding model in \texttt{BRAHMA}) of $\gtrsim 0.1$ cMpc$^{-1}$. However, in contrast to \texttt{BRAHMA}, we see accretion onto the MBH being the dominant growth mechanism. This is likely because \texttt{BRAHMA} seeds their BHs and stars at comparatively lower densities of 0.1 cm$^{-3}$ compared to our 100 cm$^{-3}$ (see Figure \ref{fig:phase}), meaning that gas is converted into stars/BHs before it can become dense and cold enough to be accreted efficiently. These objects then merge together instead of being accreted directly. Additionally, while our stellar cluster particles contain a range of stellar masses and lifetimes, resulting in staggered SNe explosions, stellar clusters in \texttt{BRAHMA} are modelled as single stellar species i.e. they all explode simultaneously, providing very strong feedback and preventing accretion. On the other hand, at higher spatial and mass resolution, the \texttt{SEEDZ} framework will likely also see a larger contribution from mergers to the growth rate.\\
\indent The \texttt{MELI$\odot$RA} simulations \citep{Cenci2025} use a MBH seeding model more similar to that of \texttt{SEEDZ}, by specifically using a mass inflow rate as one of their criteria. Overall the \texttt{MELI$\odot$RA} simulations are somewhat coarser than \texttt{SEEDZ} (by a factor of a few) but are able to evolve to significantly lower redshift ($z_{f\rm inal} = 4$). For example, \cite{Cenci2025} specifically target the redshift evolution of LRDs, as that comes within their predictive sphere. In contrast to \texttt{SEEDZ} and similar to \texttt{BRAHMA}, \texttt{MELI$\odot$RA} uses an effective equation of state for gas cooling which means that the cold gas phase is again missed, having knock-on effects on MBH formation and accretion. However, similar to our results they find that MBH experience rapid early feedback, which decreases accretion within 200 Myr in their case (in our case this occurs more rapidly - a few tens of Myr).

\section{Conclusions}
\label{sec:conclusions}
We have introduced the \texttt{SEEDZ} simulations, a suite of cosmological hydrodynamic simulations exploring the formation and growth of the first massive black holes (MBHs). The goal of our simulations is to model realistic formation and growth channels for seed black holes. We do this by modelling both the formation of light and heavy seeds in a self-consistent way within the simulation setup. Light seeds form from the end points of PopIII stars while heavy seeds form in regions undergoing rapid gravitational collapse similar to the conditions necessary for the formation of supermassive stars, or when fragmentation is too intense to allow supermassive star formation, the formation of a dense stellar cluster. At our current resolution we cannot yet follow the growth of the light seeds accurately but this is a current limitation we are working to address. \\
\indent The simulations were performed with the \textsc{arepo2} code, utilising our newly developed and physically motivated BH seeding mechanisms. The new additions to the code include Population III star formation, supernovae (SNe) explosions (and the resulting formation of light seed BHs), metal enrichment and subsequent Population II star formation, heavy seed BH formation and Eddington/super-Eddington accretion feedback schemes. Here, we have presented our current results at $z=15$, revealing emergent behaviour of the MBH  population:

\begin{itemize}\setlength\itemsep{0.2em} 
    \item Black holes initially grow faster than their host galaxy and hence over-massive black holes are a feature of the high-z Universe.

    \item The fundamental black hole galaxy relationships we observe at $z = 0$ (most especially the $\rm{M_{BH} - M_*}$ relationship) likely only emerge in mature galaxies. At high-redshift that relationship has not yet been established.

    \item We find that already by $z=15$, MBHs can grow from their initial heavy seed masses (ranging from $5 \times 10^3 \ \rm{M}_{\odot}- 10^5$ \msolarc) up to 10$^6$ \msolarc - an increase in their initial seed masses by more than two orders of magnitude.

    \item Our MBHs grow initially at more ten times the Eddington rate for $\sim$1-15 Myr before the accretion rate falls below the Eddington rate. We have not yet observed in the simulations any retriggering of rapid growth. 

    \item The MBH growth we have captured down to $z = 15$ is accretion dominated ($>99$\%) and we see little merger driven growth ($<1$\%). However, the lack of merger driven growth is likely due to under-resolving the dynamics of MBHBs. 
    
    \item Although our heavy seeding criteria is metallicity independent, we find that the majority of our MBHs were formed in low metallicity environments. While the metallicty surrounding most of these growing MBHs increased over time, a number of MBHs remain in low metallicity environments until the current results at $z=15$.
\end{itemize}

Our results must however be interpreted in the light of their resolution. Our dark matter particle mass is $\rm{M_{DM}} \sim 7 \times 10^4$ \msolar and our minimum cell size is approximately 6 (physical) pc at $z = 15$, which while highly resolved is not yet adequate to capture the growth of light seed black holes and moreover the dynamics of our lower mass heavy seeds are only marginally resolved. Hence, we likely miss a large fraction of MBH mergers (the growth rates may in fact be underestimated in our simulations). However, some trends are clear
- in particular the emergence of an over-massive black hole population at high-z followed by galaxy growth bringing the overall M$_{\rm BH}$ - M$_*$ into line with observed relationships.

 The simulations will continue to run down to $z=10$ where will will perform a comprehensive comparison of MBH number densities and BH-to-stellar mass ratios with JWST observations (Mehta et al. in prep). These simulations are a first step towards a self-consistent approach to modelling black hole evolution in the high redshift Universe.

\section*{Acknowledgements}
\noindent JR \& JB acknowledges support from the Royal Society and Research Ireland under grant number 
 URF\textbackslash R1\textbackslash 191132. LP, DM \& JR acknowledge support from the Research Ireland Laureate programme under grant number IRCLA/2022/1165. JHW acknowledges support from NSF grants AST-2108020 and AST-2510197 and NASA grant 80NSSC21K1053. RSB acknowledges support from a UKRI Future Leaders Fellowship MR/Y015517/1.
 \ \
The simulations were performed on the Luxembourg national supercomputer MeluXina and the Czech Republic EuroHPC machine Karolina.
The authors gratefully acknowledge the LuxProvide teams for their expert support.
\ \ 
The authors wish to acknowledge the Irish Centre for High-End Computing (ICHEC) for the provision of computational facilities and support.
\ \
The authors acknowledge the EuroHPC Joint Undertaking for awarding this project access to the EuroHPC supercomputer Karolina, hosted by IT4Innovations through a EuroHPC Regular Access call (EHPC-REG-2023R03-103) and to the LuxProvide supercomputer Meluxina  through a EuroHPC Regular Access call (EHPC-REG-2025R01-008).
\ \
SK has been supported by a Research Fellowship from the Royal Commission for the Exhibition of 1851.
\ \ 
MAB is supported by a UKRI Stephen Hawking Fellowship (EP/X04257X/1).
\ \ 
DS acknowledges support from the Science and Technology Facilities Council (STFC) under grant ST/W000997/1.
\ \
RSK acknowledges financial support from the European Research Council via the ERC Synergy Grant ``ECOGAL'' (project ID 855130),  from the German Excellence Strategy via the Heidelberg Cluster of Excellence (EXC 2181 - 390900948) ``STRUCTURES'', from the German Science Foundation under grant KL 1358/22-1, and from the German Federal Ministry for Economic Affairs and Energy in project ``MAINN'' (funding ID 50OO2206). RSK is grateful for computing resources provided by the Ministry of Science, Research and the Arts (MWK) of the State of Baden-W\"{u}rttemberg through bwHPC and the German Science Foundation (DFG) through grants INST 35/1134-1 FUGG and 35/1597-1 FUGG, and also for data storage at SDS@hd funded through grants INST 35/1314-1 FUGG and INST 35/1503-1 FUGG. RSK also thanks for computing time provided by the Leibniz Rechenzentrum via grants pr32lo, pr73fi and GCS large-scale project 10391. 
\ \ 
PCC acknowledges support from a STFC Small Award (UKRI1187): ``Probing the origins of stars and life with a new approach to chemical modelling''.
\ \
MGH has been supported by STFC consolidated grants ST/N000927/1 and ST/S000623/1.

\bibliographystyle{aasjournal}
\bibliography{references_seedz}



\end{document}